\documentclass[aps,prl,twocolumn,preprintnumbers]{revtex4}
\usepackage{graphicx}
\usepackage{amssymb}
\usepackage{amsmath}
\newcommand{\be}{\begin{equation}}
\newcommand{\ee}{\end{equation}}
\newcommand{\bea}{\begin{eqnarray}}
\newcommand{\eea}{\end{eqnarray}}
\newcommand{\Eq}[1]{Eq.\,(\ref{#1})}
\newcommand{\Fig}[1]{Fig.\,\ref{#1}}
\newcommand{\Onlinecite}[1]{Ref.\,\cite{#1}} 

\newcommand{\bmu}{\mbox{\boldmath ${\mu}$}}
\newcommand{\GF}{\hat{\mathbf{G}}}

\newcommand{\br}{\mathbf{r}}

\newcommand{\En}{\mathbf{E}_n}

\newcommand{\heps}{\hat{\boldsymbol{\epsilon}}}
\newcommand{\hmu}{\hat{\boldsymbol{\mu}}}

\begin{document}
\title{Exact Mode Volume and Purcell Factor of Open Optical Systems}
\author{E.\,A. Muljarov}
\author{W. Langbein}
\affiliation{School of Physics and Astronomy, Cardiff University, Cardiff CF24 3AA,
United Kingdom}
\begin{abstract}
The Purcell factor quantifies the change of the radiative decay of a dipole in an electromagnetic
environment relative to free space. Designing this factor is at the heart of  photonics technology,
striving to develop ever smaller or less lossy optical resonators. The Purcell factor can be expressed
using the electromagnetic eigenmodes of the resonators, introducing the notion of a mode volume for
each mode. This approach allows to use an analytic treatment, consisting only of sums over eigenmode
resonances, a so-called spectral representation. We show in the present work that the expressions for
the mode volumes known and used in literature are only approximately valid for modes of high quality
factor, while in general they are incorrect. We rectify this issue, introducing the exact normalization
of modes. We present an analytic theory of the Purcell effect based on the exact mode normalization and
resulting effective mode volume. We use a homogeneous dielectric sphere in vacuum, which is
analytically solvable, to exemplify these findings.
\end{abstract}
\date{\today}
\maketitle

In his short communication~\cite{PurcellPR46} published in 1946, E.\,M. Purcell introduced a factor of
enhancement of the spontaneous emission rate of a dipole of frequency $\omega$ resonantly coupled to a
mode in an optical resonator, which is now known as the Purcell factor (PF). He estimated this factor
as
\be
F= \frac{6\pi c^3 Q_n}{\omega^3 V_n},
\label{PF0}
\ee
with the speed of light $c$, the quality factor $Q_n$ of the optical mode $n$, and its effective volume
$V_n$, the latter being evaluated
as simply the volume of the resonator. This rough estimate of $V_n$ has subsequently been refined
\cite{CoccioliIEE98,foot1} to
\be
\frac{1}{V_n}=[{\bf e}\cdot \En(\br_d)]^{2}\,,
\label{Volume}
\ee
where $\br_d$ is the position of the dipole and ${\bf e}$ the unit vector of its polarization. In this
expression, the electric field of the mode ${\bf E}_n (\br)$ is normalized \cite{CoccioliIEE98} as
\be
1 =\int_{\cal V}\varepsilon(\br) {\bf E}_n^2 (\br) d{\bf r}\,,
\label{norm1}
\ee
where $\varepsilon(\br)$ is the permittivity of the resonator. The integration is performed over the
``quantization volume'' ${\cal V}$.  However, for an open system this volume is not defined, and simply
extending ${\cal V}$ over the entire space leads to a diverging normalization integral since eigenmodes
of an open system grow exponentially outside of the system due to their leakage. This issue was mostly
ignored in the literature and patched by phenomenologically choosing a finite integration volume. Such
an approach can result in relatively small errors when dealing with modes of high $Q_n$, as we will see
later. However, the fundamental problem of calculating the exact mode normalization and thus of the
mode volume remained.

Recently, a solution to this problem has been suggested. Kristensen {\it et
al}.~\cite{KristensenOL12,KristensenACS14} have used the normalization which was introduced by Leung
{\it et al.}~\cite{LeungPRA94} for one-dimensional (1D) optical systems and later
applied~\cite{LeungJOSAB96} to three dimensions. In this approach, the volume integral in \Eq{norm1} is
complemented by a surface term and the limit of infinite volume ${\cal V}$ is taken:
\be
1=\lim_{{\cal V}\to\infty} \int_{{\cal V}} \varepsilon(\br) {\bf E}_n^2 (\br) d{\bf r} +
\frac{ic}{2\omega_n} \oint_{S_{\cal V}} {\bf E}_n^2 (\br) dS\,,
\label{norm2}
\ee
where $\omega_n$ is the mode eigenfrequency and $S_{\cal V}$ is the boundary of ${\cal V}$. It was
numerically found \cite{KristensenOL12} that the surface term was leading to a stable value of the
integral for the rather small volumes available in two-dimensional (2D) finite difference in time domain (FDTD)
calculations. However, it was noted that this was not the case for low-Q modes. We show later that \Eq{norm2}, which we call in the following the Leung-Kristensen (LK) normalization, is actually diverging in the limit ${\cal V}\to\infty$,
and therefore the generalization in \cite{LeungJOSAB96} and the LK normalization are incorrect.
Therefore, while being a cornerstone of the theory of open systems and, in particular, of the
electromagnetic theory, a correct normalization of modes for determining the mode volume in \Eq{Volume}
and thus the PF was not available in the literature.

The normalization of eigenstates is at the heart of any perturbation theory, and its absence for open
systems explains also the historical fact that an exact perturbation theory was unavailable in
electromagnetics until recently. Only in 2010 such a theory, the resonant-state expansion (RSE),
was formulated~\cite{MuljarovEPL10} and subsequently applied to 1D, 2D and 3D
systems~\cite{MuljarovEPL10,DoostPRA12,DoostPRA13,ArmitagePRA14,DoostPRA14}, demonstrating its ability
to accurately and efficiently calculate resonant states (RSs) -- the eigenmodes  -- of a perturbed open
optical system using the spectrum of RSs of a simpler, unperturbed one. The normalization of RSs
introduced in~\cite{MuljarovEPL10} is a key element of the RSE, and paves the way for an exact
calculation of mode volume and PF in optical resonators.

Here, we present a rigorous theory of the Purcell effect, based on a general exact formula for the mode
volume in arbitrary optical systems, and illustrate it on the exactly solvable model of a dielectric
spherical resonator.

In the weak coupling regime, the spontaneous emission rate of a quantum dipole, which is determining
the local density of states and the spectral function of the resonator, has the following
form~\cite{GlauberPRA91,DungPRA00,DungPRA01}, as detailed in the Supplemental Material~\cite{SM}:
\be
\gamma(\omega)=-\frac{\omega^2}{\varepsilon_0\hbar c^2}\bmu\cdot{\rm
Im}\,\GF(\br_d,\br_d;\omega)\bmu\,,
\label{SD}
\ee
where $\bmu= \mu {\bf e}$ is the electric dipole moment and $\varepsilon_0$ is the vacuum permittivity.
The dyadic Green's function (GF) $\GF$
which contributes to \Eq{SD} respects the outgoing wave boundary conditions and satisfies Maxwell's
wave equation with a delta function source term,
 \be
\left(\frac{\omega^2}{c^2}\heps(\br)- \nabla\times\nabla\times
\right)\GF(\br,\br';\omega)=\hat{\mathbf{1}}\delta(\br-\br')\,,
 \label{GFequ}
\ee
where $\heps(\br)$ is the dielectric tensor of the open optical system and $\hat{\mathbf{1}}$ is the
unit tensor. The permeability is assumed to be $\hmu(\br)=\hat{\mathbf{1}}$ throughout this paper. With
modern electromagnetic software, \Eq{GFequ}
can be solved numerically by replacing the $\delta$-like source term with a finite-size dipole. The
mode volume can then be evaluated by calculating numerically the residues of the GF at its poles, as
has been recently shown~\cite{BaiOE13}. Such a fully numerical approach circumvents the definition of
the mode volume in terms of the mode field.
In an alternative method introduced by C. Sauvan {\it et al.}~\cite{SauvanPRL13,AgioNPho13} the mode
volume is determined from the mode field calculated numerically including an artificial perfectly
matched layer (PML), which is widely used in electromagnetic software packages. A PML is an absorbing
layer which allows to efficiently simulate outgoing boundary conditions within a finite simulation
volume. The divergence of the normalization \Eq{norm1} is avoided by converting the radiative losses to
the outside region into absorptive losses within the simulation volume. For the example provided in
Ref.~\cite{SauvanPRL13} of a mode in a 100\,nm diameter gold sphere we found good agreement of the
numerical value of the mode volume with one obtained by the exact normalization method introduced in
the present work, as detailed  in Table~S1 in \Onlinecite{SM}.

The purpose of this Letter is to rectify and complete the Purcell theory by providing the exact general
formulas for the mode normalization, mode volume, and resulting PF. The normalization deals with only
the mode field in a finite volume and its frequency, so that modes calculated by any available means
can be used.

Inside the optical system, i.e. within the volume of inhomogeneity of $\heps(\br)$, the GF has the
following spectral representation~\cite{MuljarovEPL10,DoostPRA13,DoostPRA14}
\be
\GF(\br,\br';\omega)=c^2\sum _n\frac{\En(\br)\otimes\En(\br')}{2\omega_n(\omega-\omega_n)}\,,
\label{ML}
\ee
in which the sum is taken over all RSs. These are the optical modes of the system, the eigen-solutions
of Maxwell's wave equation satisfying the outgoing wave boundary conditions. The eigenfrequency
$\omega_n=\Omega_n-i\Gamma_n$ of the RS is generally complex and contains the position $\Omega_n$ of
the resonance and its half width at half maximum $\Gamma_n$. The quality factor of each RS is given by
$\displaystyle Q_n=\Omega_n/2\Gamma_n$. The spectral representation \Eq{ML} requires that the RSs (with
$\omega_n\neq 0$) are normalized according to
\bea
&&1=\int_{{\cal V}}\En (\br)\cdot\heps(\br)\En(\br)\,d{\bf r}
\label{norm}
\\
&&+\frac{c^2}{2\omega^2_n}\oint_{S_{\cal V}} \left[\En\cdot\frac{\partial}{\partial
s}(\br\cdot \nabla)\En- \frac{\partial \En}{\partial s} \cdot (\br\cdot \nabla)\En\right] dS \nonumber
\eea
where the first integral is taken over an arbitrary simply connected volume ${\cal V}$ enclosing the
inhomogeneity of the system, while the second integral
is taken over a closed surface  $S_{\cal V}$, the boundary of ${\cal V}$, with the normal derivative $\partial/\partial s=\hat{\bf n}\cdot \nabla$
using the surface normal $\hat{\bf n}$.
Equation (\ref{norm}) is the {\it correct mode normalization},
compatible with the spectral representation \Eq{ML} of the GF. The volume ${\cal V}$ can be arbitrarily big -- both integrals in \Eq{norm} grow
exponentially with ${\cal V}$ but exactly compensate each other, making the result independent of ${\cal
V}$. We emphasize that the normalization  \Eq{norm} is valid for {\it any} surface $S_{\cal V}$ of
integration outside the system, and does not require taking the limit ${\cal V}\to \infty$, in contrast to the LK normalization.

The expression in the surface term of \Eq{norm} can be simplified in spherical coordinates to radial derivatives, using
$\br\cdot \nabla= r\partial/\partial r$. Furthermore, choosing ${\cal V}$ in the form of a finite sphere in 3D or a finite
cylinder in 2D yields $\partial /\partial s=\partial /\partial r$ and a simpler form of the
normalization of RSs~\cite{MuljarovEPL10,DoostPRA13,DoostPRA14}. A proof of the normalization \Eq{norm} using a spherical volume ${\cal V}$ is given in
Ref.~\cite{DoostPRA14}. Since a convenient normalization volume ${\cal V}$ can be different from a
sphere, we have generalized here the normalization to an arbitrarily shaped simply connected volume and have presented it in the form independent of the coordinate system used. The related proof of \Eq{norm} is provided in \Onlinecite{SM}. In the presence of a frequency dispersion of the
permittivity, which is important e.g. in metallic resonators, the dielectric constant $\heps(\br)$ in \Eq{norm} is replaced by $\displaystyle
\partial\left(\omega^2\heps(\br,\omega)\right)/\partial \left(\omega^2\right)$, as shown in \Onlinecite{SM}, and in case of the dispersion also of the background material (replacing vacuum treated above),
the surface term in  \Eq{norm} acquires an additional factor $\displaystyle
\varepsilon^{-1}(\omega)\partial\left(\omega^2\varepsilon(\omega)\right)/\partial \left(\omega^2\right)$.

Comparing the LK and the exact normalization, we note that \Eq{norm2} has an additional prefactor $\omega_n/c$, the wavevector, while \Eq{norm} calculates explicitly the normal derivatives of the fields, which results in the factor $\omega_n/c$ only for fields propagating normal to the surface of integration. This observation clarifies the qualitative difference between the two normalizations -- the LK normalization assumes normal propagation, while the exact normalization takes into account the actual propagation direction. The exact normalization in the form of \Eq{norm} depends on both 1st and 2nd spatial derivatives of the fields. However, as shown in \Onlinecite{SM}, one can use Maxwell's equations to convert \Eq{norm} into a form containing only first derivatives. This can be advantageous for application to RSs calculated using numerical solvers such as the finite element method (FEM), as demonstrated in \Onlinecite{SM}. We find that \Eq{norm} is robust against the choice of grids and PML thickness of the FEM. Furthermore, since it can be evaluated close to the system, the surface area can be minimized, reducing the numerical error and the required simulation domain. The LK normalization instead not only diverges for large $R$, as shown before, but also has significant errors for small $R$ of the order of the system size, where the non-normal field propagation results in an error scaling as $R^{-2}$. Therefore, even for RSs of high Q for which the divergence with $R$ is slow, the LK normalization requires to use a simulation domain orders of magnitude larger than the system size, which is computationally costly.

The spectral representation \Eq{ML} determines the exact PF, taking into account the contribution of all significant modes. Indeed, using  Eqs.~(\ref{SD}) and (\ref{ML}), the PF in the weak coupling regime is obtained as
\be
F(\omega)=\frac{\gamma(\omega)}{\gamma_0(\omega)} = \frac{3\pi c^3}{\omega} \sum_n {\rm
Im}\,\frac{1}{V_n\omega_n (\omega_n-\omega)}\,,
\label{PF}
\ee
which requires using the {\it exact mode volume} $V_n$ given by \Eq{Volume} with the electric field $\En(\br)$ of the RS respecting the correct normalization \Eq{norm}.

Here $\gamma_0 =\omega^3 \mu^2 /(6\pi\varepsilon_0 \hbar c^3)$  is the radiative decay rate of the
dipole in free space~\cite{PurcellPR46}, which can be deduced from \Eq{SD} using the GF of empty
space~\cite{LevineCPAM50}, as demonstrated in \Onlinecite{SM}.
If a single mode $n$ dominates in \Eq{PF}, the PF on resonance ($\omega={\rm Re}\, \omega_n$) can be
approximated as  $F(\omega)\approx  6\pi c^3 Q_n/[\omega^2{\rm Re}(\omega_n V_n)]$. For a high-Q mode,
the eigenfrequency and mode volume are approximately real, and the latter formula reproduces Purcell's
result \Eq{PF0} when using the correct mode volume $V_n$.

\begin{figure}[t]
\includegraphics*[width=\columnwidth]{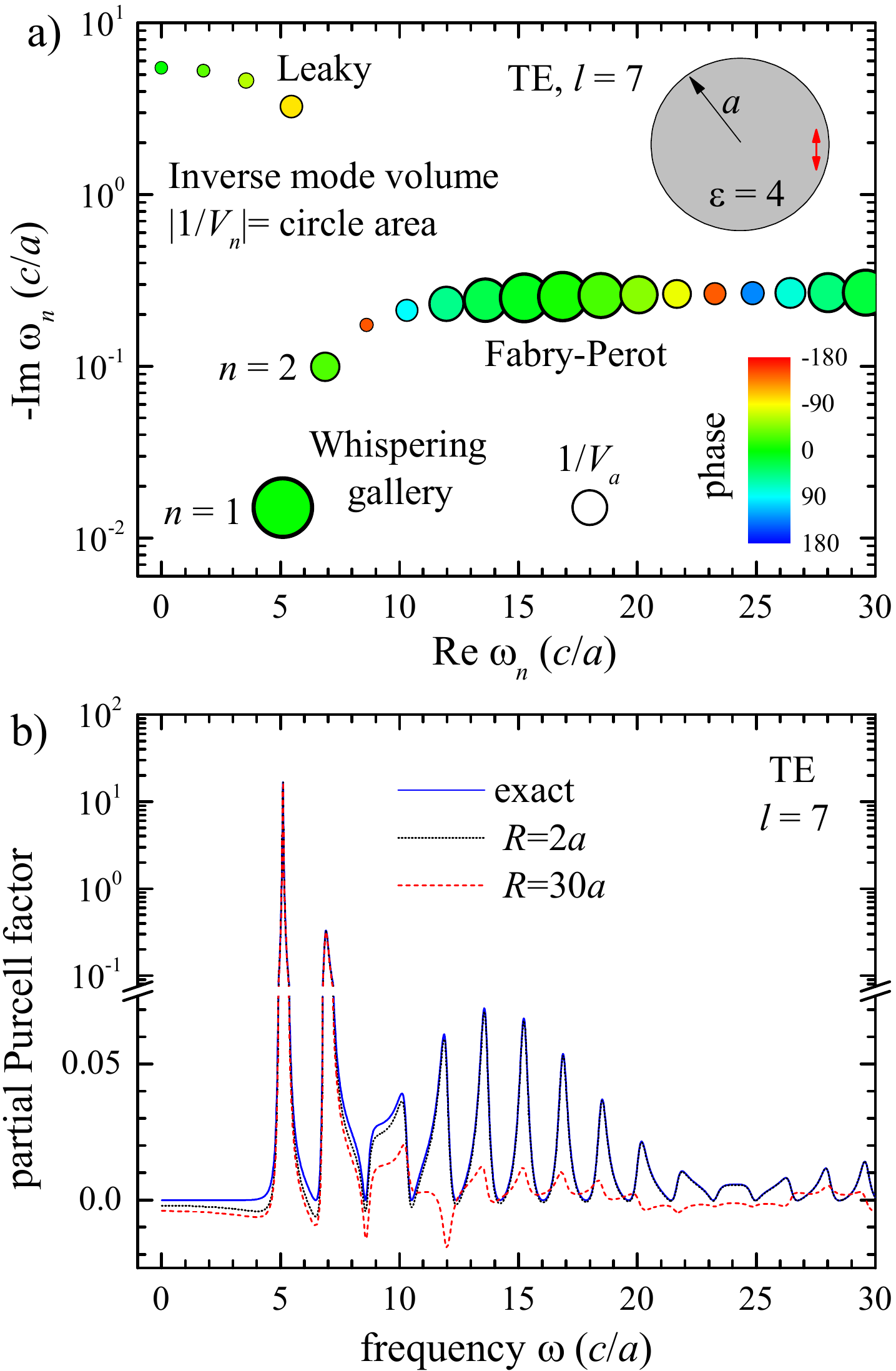}
\caption{(color online) a) Mode volumes for a dielectric sphere in vacuum, with permittivity
$\varepsilon=4$ and radius $a$, for $l=7$ TE modes and a point dipole placed at $|\br_d|=0.9a$ with
direction ${\bf e}=(0,0,1)$ in spherical coordinates (see sketch). The mode volume is presented as the
sum of the inverse mode volume over all degenerate states $m=-l,\dots,l$. Its amplitude is shown by the
circle area and its phase by the color. The volume of the sphere $V_a=4\pi a^3/3$ is shown for
comparison. The position of the circles in the complex frequency plane is given by the mode
eigenfrequency $\omega_n$.
b) Partial Purcell factor calculated via \Eq{PF} for $l=7$ and TE polarization, for the geometry of a), using the exact mode normalization (blue line) and
the LK normalization evaluated for integration volumes given by a sphere of radius $R=2a$ and
$R=30a$, as labeled.
}\label{fig:F1}
\end{figure}

\begin{figure}[t]
\includegraphics*[width=\columnwidth]{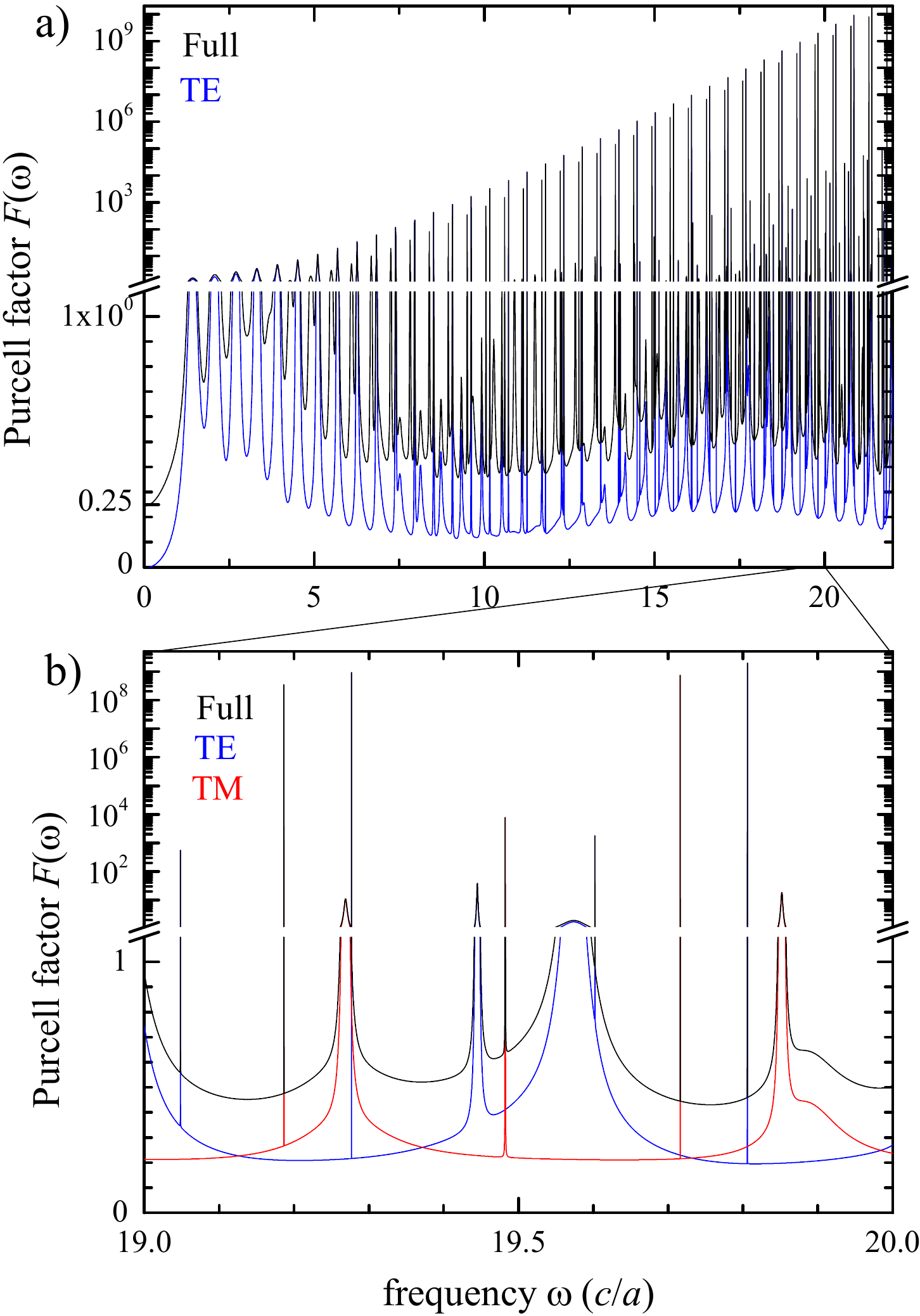}
\caption{(color online) Purcell factor versus transition frequency $\omega$ calculated by \Eq{PF} using
modes with $l<38$ and $|\omega_n|<40c/a$. The dipole is placed at $|\br_d|=0.9a$ as in \Fig{fig:F1},
and the PF is averaged over the polarization directions. a) Full PF (black line) and partial PF for TE
modes only (blue line). b) zoom of a), showing additionally the partial PF for TM modes (red line).}
\label{fig:F2}
\end{figure}

For illustration, we have calculated the mode volume and PF of a dielectric spherical resonator with
homogeneous permittivity ($\varepsilon =4$) surrounded by vacuum, for a point dipole placed at
$|\br_d|=0.9a$ with direction ${\bf e}=(0,0,1)$ in spherical coordinates (see sketch in \Fig{fig:F1}a).
The inverse mode volume of several eigenmodes with the angular momentum $l=7$ and transverse electric
(TE) polarization, summed over the degenerate states with azimuthal number $m=-l,\dots,l$, is shown in
\Fig{fig:F1}a. The modes can be classified as leaky modes, whispering gallery modes (WGMs), and
Fabry-P{\'e}rot (FP) modes, as indicated. The chosen dipole position is close to the field maximum of
the fundamental WGM, therefore its mode volume is small and essentially real. With increasing mode
order going into the FP modes, the mode volume oscillates as the field maxima and minima move across
the dipole position. Interestingly, the phase of the mode volume rotates accordingly, yielding negative
mode volumes at the positions of the field minima, at which the mode field is imaginary. This also
elucidates that the radiative decay into the modes is not a simple superposition of Lorentzian lines
describing independent channels, but shows interference. This can actually be expected, as modes of
equal $l$, $m$, and polarization couple into the same outgoing loss channel.

The resulting partial PF for $l=7$ TE modes (see \Fig{fig:F1}b) is dominated by the $n=1$ WGM providing on resonance a PF of about 20. The complex mode volume leads to non-Lorentzian features in the spectrum, due to the mode interference. The total contribution to the PF of all modes for each loss channel (for spherical symmetry all modes with equal $m$, $l$, and polarization) is strictly positive, as expected. To exemplify the issues with the LK normalization, we show the resulting PF for two finite integration volumes given by spheres of radius $R=2a$ and $R=30a$. The observed deviation, which is increasing with $R$, is showing an underestimation of the contribution of leaky and FP modes. The PF also shows negative values which are unphysical. Taking the limit $R\to\infty$, the mode volume diverges exponentially, according to \Eq{norm2}, see Fig.~S1 of \Onlinecite{SM}, -- this is also true for high-Q modes but commences at larger $R$ -- so that the PF vanishes. In metallic resonators the modes have generally low Q-factors, yielding a fast exponential growth (see
Fig.~S2) and large errors of \Eq{norm2} for any $\cal V$.

To explicitly show the validity of the spectral representation and the resulting PF based on the exact normalization, we compare it with the PF calculated using the analytic GF of a sphere. We take into account both TE and transverse magnetic (TM) polarizations and sum over all significant values of $l$ and $m$. Examples of mode volumes and partial PFs for the TM modes are shown in Figs.~S3 and S4 of \Onlinecite{SM}  for two different directions of the dipole. The resulting PF for a dipole at a distance $0.9a$ from the center of the sphere averaged over its polarization directions is shown in \Fig{fig:F2}, with partial PFs shown separately for TE modes in Figs.~2(a) and 2(b) and for TM modes in Fig.~2(b). In the low-frequency limit the well known static value of the field reduction by a factor of $3/(2+\varepsilon)$ inside a dielectric sphere in vacuum is reproduced, so that $F(0)=0.25$. Comparing these results with the ones obtained using the analytic GF of a sphere shows excellent agreement, as detailed in Sec.\,VI of \Onlinecite{SM}. The spectral zoom in \Fig{fig:F2}(b) allows to see the extremely sharp WGM lines on top of much wider resonances and their separation into TE and TM modes. Purcell factors up to $10^{10}$ are found in resonance to WGM of similarly high Q-factors.

In conclusion, we have provided a general exact analytic form of the normalization of eigenmodes in an arbitrary open optical resonator, which rectifies a normalization expression previously believed to be valid. We have shown that the normalization can be used with high accuracy also for RSs determined by numerical solvers, while at the same time the flexible normalization volume does not constrain the size of the computational domain. The correct normalization is of key importance for the electromagnetic theory, as it determines the spectral representation of the dyadic Green's function of Maxwell's wave equation, which can be used for calculation of any observable, such as scattering and extinction cross-sections and local density of states. We focussed in the present work on the consequences for the determination of the mode volume and the Purcell factor, which is a cornerstone for cavity quantum electrodynamics and nanoplasmonics. Further fundamental developments enabled by this exact normalization are to be expected, as exemplified by the rigorous perturbation theory in electrodynamics called the resonant-state expansion ~\cite{MuljarovEPL10,DoostPRA12,DoostPRA13,ArmitagePRA14,DoostPRA14}.

\acknowledgments This work was supported by the Cardiff University EPSRC Impact Acceleration Account EP/K503988/1, and the S\^er Cymru National Research Network in Advanced Engineering and Materials. The authors acknowledge discussions with T.\,Weiss and  M.\,B. Doost, and thank G.\,Zorinyants for performing the calculations for the RSs of a non-spherical system using a numerical solver.


\begin{thebibliography}{10}
\bibitem{foot1}
For consistency with results of the present paper, we have removed from the
  original formula the factor of $\varepsilon({\bf r}_d)$ and added the dipole
  polarization vector ${\bf e}$  .
\end{thebibliography}

\end{document}


\title{Exact Mode Volume and Purcell Factor of Open Optical Systems\\[6pt] Supplemental Material}
\author{E.\,A. Muljarov}
\author{W. Langbein}
\affiliation{School of Physics and Astronomy, Cardiff University, Cardiff CF24 3AA,
United Kingdom}
\begin{abstract}
\end{abstract}
\pacs{03.50.De, 42.25.-p, 03.65.Nk}
\date{\today}
\maketitle

\section{Spontaneous emission of a quantum dipole in an arbitrary dielectric system}
\label{SM-I}

The full Hamiltonian describing a quantum dipole coupled to photon states in an arbitrary open optical system is given \cite{GlauberPRA91} by
\be
H=H_0+V\,,
\tag{S\theequation}\label{Hfull}\stepcounter{equation}
\ee
where
\be
H_0=\sum_k\hbar\omega_k a_k^\dagger a_k+\hbar\omega_d d^\dagger d
\tag{S\theequation\label{H0}}\stepcounter{equation}
\ee
is the non-interacting part and
\be
V=-i\sum_k (\varphi_k d^\dagger a_k - \varphi_k^\ast a_k^\dagger d)
\tag{S\theequation}\label{interaction}\stepcounter{equation}
\ee
is the interaction between the dipole and photons in the rotating wave approximation. Here $a^\dagger_k$ is the photon creation operator in state $k$, $d^\dagger$ is the fermionic creation operator for the two-level system of the quantum dipole, with $\omega_d$ being the ground-to-excited state transition frequency, and
\be
\varphi_k=\sqrt{\frac{\hbar\omega_k}{2\varepsilon_0}} \bmu\cdot {\bf f}_k(\br_d)
\tag{S\theequation}\label{phi-equ}\stepcounter{equation}
\ee
is the coupling matrix element, in which $\bmu$ is the electric dipole moment of the point dipole placed at $\br=\br_d$, $\varepsilon_0$ is the vacuum permittivity and ${\bf f}_k(\br)$ is a vector eigenfunction of the electric field of the continuum state $k$ satisfying the Maxwell wave equation
\be
- \nabla\times\nabla\times {\bf f}_k(\br)+\frac{\omega_k^2}{c^2}\heps(\br){\bf f}_k(\br)=0
\tag{S\theequation}\label{fk}\stepcounter{equation}
\ee
with a {\it real} eigen-frequency $\omega_k\geqslant 0$. The symmetric tensor $\heps(\br)$ of the dielectric constant describes the open optical system under study and for simplicity is assumed here frequency-independent.

Following Glauber~\cite{GlauberPRA91}, we consider the Schr\"odinger equation describing the full system
($\hbar=1$ is used below for brevity of notations),
\be
i\frac{d}{dt}|\Phi(t)\rangle=H|\Phi(t)\rangle\,,
\tag{S\theequation}\stepcounter{equation}
\ee
and take its formal solution in the form
\be
|\Phi(t)\rangle=e^{-iHt} |\Phi(0)\rangle\,,
\tag{S\theequation}\stepcounter{equation}
\ee
where $|\Phi(t)\rangle$ is the wave function of the dipole-photon system. We are interested in the probability for the dipole to stay in the excited state and calculate the probability amplitude in the following way:
\be
\alpha(t)=\langle\Phi(0)|\Phi(t)\rangle = \langle 0|d e^{-iHt} d^\dagger|0\rangle=  \langle 0|d(t) U(t) d^\dagger |0\rangle\,.
\tag{S\theequation}\stepcounter{equation}
\ee
In the above equation, the dipole moment operator is written in the interaction representation, $d(t)=e^{iH_0t}d e^{-iH_0t}$. We have also assumed  that in the initial state, the photon subsystem is in its ground state and the quantum dipole is in its excited state, i.e.
$|\Phi(0)\rangle =d^\dagger|0\rangle$, where $|0\rangle$ is the ground state of the full system. The evolution operator $U(t)=e^{iH_0t} e^{-iHt}$ satisfies the equation
\be
i \frac{d U(t)}{dt}= V(t) U(t)\,,
\tag{S\theequation}\stepcounter{equation}
\ee
where $V(t)=e^{iH_0t}V e^{-iH_0t}$. Its solution can be written as an infinite perturbation series
\begin{align}
U(t)=&\ 1+(-i) \int_0^t V(t_1) d t_1 \nonumber \\
&+(-i)^2\int_0^t V(t_1) d t_1\int_0^{t_1} V(t_2) d t_2 + \dots
\tag{S\theequation}\stepcounter{equation}
\end{align}
To calculate $\alpha(t)$, we evaluate
\begin{align}
&V(t_1)V(t_2) d^\dagger|0\rangle=-e^{iH_0t_1}\sum_{k_1}(\varphi_{k_1} d^\dagger a_{k_1} - \varphi_{k_1}^\ast a_{k_1}^\dagger d) \nonumber\\
&\times e^{-iH_0(t_1-t_2)}\sum_{k_2}(\varphi_{k_2} d^\dagger a_{k_2} - \varphi_{k_2}^\ast a_{k_2}^\dagger d)e^{-iH_0t_2}d^\dagger|0\rangle\nonumber\\
&=e^{i\omega_d(t_1-t_2)}\sum_k |\varphi_k|^2 e^{-i\omega_k(t_1-t_2)}d^\dagger|0\rangle\,.
\tag{S\theequation}\stepcounter{equation}
\end{align}
Then for $t>0$, $\alpha(t)$ can be written in the form of an integral equation:
\begin{align}
\alpha(t)=&\ e^{-i\omega_d t}- \sum_k |\varphi_k|^2  \int_0^t d t_1\int_0^{t_1} d t_2
\nonumber\\ &
\times e^{-i\omega_d(t-t_1)} e^{-i\omega_k(t_1-t_2)}\alpha(t_2)
\tag{S\theequation}\stepcounter{equation}
\end{align}
which can be solved explicitly in the Fourier space:
\be
\tilde{\alpha}(\omega)=\frac{-i}{\omega-\omega_d-\Sigma(\omega)}\,,
\tag{S\theequation}\label{alpha-FT}\stepcounter{equation}
\ee
where $\tilde{\alpha}(\omega)$ is the time Fourier transform of $\alpha(t)$,  and the self-energy $\Sigma(\omega)$ is given by a formula
\be
\Sigma(\omega)=\frac{1}{\hbar^2} \sum_k \frac{|\varphi_k|^2}{\omega-\omega_k+i\delta_+}\,,
\tag{S\theequation}\label{self}\stepcounter{equation}
\ee
in which $\delta_+\to 0_+$ and $\hbar$ has been restored.  Note that the problem described by Eqs.~(\ref{Hfull})--(\ref{interaction}) is the famous exactly solvable Fano-Anderson problem. Indeed, owing to the bilinear form of the interaction
\Eq{interaction} the exact perturbation series for the self-energy ends in first order~\cite{Mahan}.

Let us express the self-energy $\Sigma(\omega)$ in terms of the dyadic GF of Maxwell's wave equation. The full time-dependent GF $\GFP(\br,\br'; t-t')$ satisfies the equation
\be
- \nabla\times\nabla\times \GFP - \frac{\heps(\br)}{c^2}\frac{\partial^2\GFP}{\partial t^2}=\hat{\mathbf{1}}\delta(\br-\br')\delta(t-t')
\tag{S\theequation}\label{GFequ1}\stepcounter{equation}
\ee
and has the following explicit form in terms of the continuum eigenstates, the solutions of \Eq{fk}:
\be
\GFP(\br,\br'; t-t')=\frac{c^2}{i}\sum_k\frac{{\bf f}_k(\br)\otimes {\bf f}^\ast_k(\br')}{2\omega_k}\,e^{-i\omega_k|t-t'|}\,.
\tag{S\theequation}\label{GFfull}\stepcounter{equation}
\ee
Note that substituting \Eq{GFfull} into \Eq{GFequ1} and using \Eq{fk} results in the closure relation for the continuum eigenstates,
\be
\heps(\br) \sum_k {\bf f}_k(\br)\otimes {\bf f}^\ast_k(\br')=\hat{\mathbf{1}}\delta(\br-\br')\,.
\tag{S\theequation}\stepcounter{equation}
\ee
Fourier transforming the GF given by \Eq{GFfull} versus $t-t'$ we obtain
\be
\GF(\br,\br';\omega)=c^2 \sum_k \frac{{\bf f}_k(\br)\otimes {\bf f}^\ast_k(\br')}{\omega^2-\omega_k^2+i\delta_+}\,.
\tag{S\theequation}\stepcounter{equation}
\ee
Then, using \Eq{phi-equ} for a positive frequency $\omega$ we find
\begin{align}
I(\br_d,\omega)\equiv&\ \bmu\cdot{\rm Im}\,\GF(\br_d,\br_d;\omega)\bmu\nonumber\\
=&\ -\frac{\pi\varepsilon_0 c^2}{\hbar}\sum_k \frac{|\varphi_k|^2}{\omega_k^2}\delta(\omega-\omega_k)\,,
\tag{S\theequation}\stepcounter{equation}
\end{align}
that allows us to express the self-energy in terms of the projection $I(\br,\omega)$ of the GF tensor:
\begin{align}
\Sigma(\omega)=&\ \frac{1}{\hbar^2} \sum_k \int_0^\infty \frac{|\varphi_k|^2 \delta(\omega'-\omega_k)d\omega'}{\omega-\omega'+i\delta_+}
\nonumber\\
=&\ -\frac{1}{\pi\varepsilon_0 c^2\hbar} \int_0^\infty \frac{I(\br_d,\omega'){\omega'}^2}{\omega-\omega'+i\delta_+} d\omega'\,.
\tag{S\theequation}\label{SE2}\stepcounter{equation}
\end{align}
For the GF of a homogeneous medium with the dielectric constant $\varepsilon$ we have
\be
{\rm Im}\,\GF^{0}(\br,\br;\omega)=-\frac{\sqrt{\varepsilon}\omega}{6\pi c} \hat{\mathbf{1}}
\tag{S\theequation}\label{GF0}\stepcounter{equation}
\ee
as shown below, such that the integral in \Eq{SE2} diverges for large $\omega'$, which is the well known divergence problem of the Lamb shift, usually treated by introducing a frequency cut-off. For an inhomogeneous open optical system this integral is, however, convergent. Indeed, using the spectral representation of the GF in terms of resonant states (RSs) with {\it complex} eigenfrequencies $\omega_n$~\cite{MuljarovEPL10,DoostPRA14}
\be
\GF(\br,\br';\omega)=c^2\sum _n\frac{\En(\br)\otimes\En(\br')}{2\omega(\omega-\omega_n)}
\tag{S\theequation}\label{GFn}\stepcounter{equation}
\ee
(see also Sec.~\ref{SM-II}), we obtain for any ${\bf r}$ inside the system
\be
I(\br,\omega)=\frac{c^2}{2\omega} {\rm Im}\sum_n \frac{g^2_n(\br)}{\omega-\omega_n}\,,
\tag{S\theequation}\label{Idef}\stepcounter{equation}
\ee
where
\be
g_n(\br)=\bmu\cdot\En(\br)\,.
\tag{S\theequation}\stepcounter{equation}
\ee
We note that RSs contribute to \Eq{GFn} in pairs: Each RS $n$ with the eigenfrequency $\omega_n$ and electric field eigenfunction $\En(\br)$ has a counterpart $-n$ with $\omega_{-n}=-\omega^\ast_n$ and ${\bf E}_{-n}(\br)=\En^\ast(\br)$. Their joint contribution to \Eq{Idef}  is given by
\begin{align}
&{\rm Im}\,\left[ \frac{g_n^2(\br)}{\omega-\omega_n}+ \frac{{g^\ast_n}^2(\br)}{\omega+\omega^\ast_n}\right]
\tag{S\theequation}\stepcounter{equation}
\\
&\ =\frac{A''_n(\omega-\omega'_n)+A'_n{\omega''_n}}{(\omega-\omega'_n)^2+{\omega''_n}^2}+
\frac{-A''_n(\omega+\omega'_n)+A'_n{\omega''_n}}{(\omega+\omega'_n)^2+{\omega''_n}^2}\,,\nonumber
\end{align}
where $g_n^2(\br)=A'_n+iA''_n$ and $\omega_n=\omega'_n+i\omega''_n$\,. Therefore $I(\br,\omega)\propto 1/\omega^3$ at $\omega\to\infty$ and the integral in \Eq{SE2} converges.

For small values of $\omega$, it is more practical to use a different form of the GF~\cite{MuljarovEPL10,DoostPRA14}
\be
\GF(\br,\br';\omega)=c^2\sum _n\frac{\En(\br)\otimes\En(\br')}{2\omega_n(\omega-\omega_n)}\,,
\tag{S\theequation}\label{GFalt}\stepcounter{equation}
\ee
which follows from \Eq{GFn} and the sum rule (see Sec.~\ref{SM-II})
\be
\sum _n\frac{\En(\br)\otimes\En(\br')}{\omega_n} =0\,.
\tag{S\theequation}\label{sumrule}\stepcounter{equation}
\ee
Then $I(\br,\omega)$ can then be written as:
\be
I(\br,\omega)=\frac{c^2}{2} {\rm Im} \sum_n\frac{g^2_n(\br)}{\omega_n(\omega-\omega_n)}\,,
\tag{S\theequation}\label{Ialt}\stepcounter{equation}
\ee
and the contribution of the pair of poles takes the form
\begin{align}
&{\rm Im}\,\left[ \frac{g^2_n(\br)}{\omega_n(\omega-\omega_n)}- \frac{{g^\ast_n}^2(\br)}{\omega^\ast_n(\omega+\omega^\ast_n)}\right]
\tag{S\theequation}\stepcounter{equation}
\\
&\ =\frac{B''_n(\omega-\omega'_n)+B'_n{\omega''_n}}{(\omega-\omega'_n)^2+{\omega''_n}^2}-
\frac{-B''_n(\omega+\omega'_n)+B'_n{\omega''_n}}{(\omega+\omega'_n)^2+{\omega''_n}^2}\,,\nonumber
\end{align}
where $g_n^2(\br)/\omega_n=B'_n+iB''_n$\,, so that  $I(\br,\omega)\propto \omega$ in the limit $\omega\to 0$.

Note that for the poles of the GF on the imaginary $\omega$-axis which do not have counterparts, $g_n^2(\br)$ is real, and both low- and high-frequency asymptotics of $I(\br,\omega)$ obtained above are preserved. Moreover, for the same reason, static modes (having $\omega_n=0$) do not contribute to the spontaneous emission, as the corresponding term of the GF is purely real (the modes are localized).

Using Eqs.~(\ref{Idef}) or (\ref{Ialt}), $\Sigma(\omega)$ can be calculated analytically for any finite number of RSs, thus providing direct access to the analytic continuation  of $\tilde{\alpha}(\omega)$ into the complex $\omega$ plane and to its pole structure. Owing to the causality principle, $\tilde{\alpha}(\omega)$ has poles only in the lower half plane, which results in the following expression for the probability amplitude in the time domain:
\be
\alpha(t)= \theta(t)\sum_j c_j e^{-i(\omega_d+\delta\omega_j)t-\gamma_j t}\,,
\tag{S\theequation}\stepcounter{equation}
\ee
where
\be
\frac{1}{c_j}=1-\left.\frac{d\Sigma(\omega)}{d\omega}\right|_{\omega=\omega_j}
\tag{S\theequation}\stepcounter{equation}
\ee
and $\omega_j= \omega_d+\delta\omega_j-i\gamma_j$ are the poles of $\tilde{\alpha}(\omega)$. Such an analysis is important for the strong coupling regime. In the weak coupling regime instead, $\Sigma(\omega)$ can be considered as a small correction, and $\tilde{\alpha}(\omega)$ can be treated in the single-pole approximation leading to
\be
\alpha(t)= \theta(t)e^{-i(\omega_d+\delta\omega)t-\gamma t}\,,
\tag{S\theequation}\label{al}\stepcounter{equation}
\ee
where $\delta\omega-i\gamma=\Sigma(\omega_d)$ is the self-energy correction to the pole of GF of the dipole, calculated ``on-shell'', i.e. at $\omega=\omega_d$. The Lamb shift $\delta\omega$ and the spontaneous emission rate $\gamma$ then take the following explicit form
\be
\delta\omega=\frac{1}{\pi\varepsilon_0 c^2\hbar}\,\dashint_0^\infty\frac{I(\br_d,\omega) \omega^2 d\omega}{\omega-\omega_d}\,,
\tag{S\theequation}\label{del}\stepcounter{equation}
\ee
\be
\gamma(\omega_d)=-\frac{\omega_d^2}{\varepsilon_0 c^2\hbar} I(\br_d,\omega_d),
\tag{S\theequation}\label{gam}\stepcounter{equation}
\ee
where the principal value integral is introduced in \Eq{del}. Equations~(\ref{al})--(\ref{gam}) are know in the literature as the Weisskopf-Wigner approximation~\cite{DungPRA00}.

Let us check that \Eq{gam} produces the correct expression for the spontaneous emission rate in the case of a homogeneous dielectric medium.
The GF of the free space satisfies the equation
\be
- \nabla\times\nabla\times \GF^0(\br,\br';\omega) + k^2\GF^0(\br,\br';\omega)=\hat{\mathbf{1}}\delta(\br-\br')\,,
\tag{S\theequation}\label{GF0equ}\stepcounter{equation}
\ee
where $k^2=\varepsilon \omega^2/c^2$ and $\varepsilon$ is the dielectric constant of the medium. The solution of \Eq{GF0equ} has the form~\cite{Levine50}:
\be
\GF^0(\br,\br';\omega)=-\left(\hat{\mathbf{1}}+\frac{1}{k^2}\nabla\otimes\nabla\right) \frac{e^{ik|\br-\br'|}}{4\pi|\br-\br'|}\,,
\tag{S\theequation}\stepcounter{equation}
\ee
or, more explicitly~\cite{Martin98},
\be
\GF^0(\br,\br';\omega)=C\, \hat{\mathbf{1}}+D\, \frac{{\bf b}\otimes{\bf b}}{b^2}\,,
\tag{S\theequation}\stepcounter{equation}
\ee
where ${\bf b}=\br-\br'$, $b=|{\bf b}|$ and
\be
C=-\left(1+\frac{ikb-1}{k^2b^2}\right)\frac{e^{ikb}}{4\pi b}= \frac{2-k^2b^2}{8\pi k^2 b^3}-i\frac{k}{6\pi}+k\mathcal{O}(kb)\,,
\tag{S\theequation}\stepcounter{equation}
\ee
\be
D=-\frac{3-3ikb-k^2b^2}{k^2b^2}\frac{e^{ikb}}{4\pi b}= -\frac{6+k^2b^2}{8\pi k^2 b^3}+k\mathcal{O}(kb)\,,
\tag{S\theequation}\stepcounter{equation}
\ee
expanded up to zeroth order in $kb$. Taking the limit $\br'\to\br$, so that $b\to0$, we obtain
\Eq{GF0} and finally
\be
\gamma_0(\omega)=-\frac{\omega^2}{\varepsilon_0\hbar c^2}\bmu\cdot{\rm Im}\,\GF^0(\br_d,\br_d;\omega)\bmu=\frac{\sqrt{\varepsilon}\omega^3\mu^2}{6\pi\epsilon_0\hbar c^3}\,,
\tag{S\theequation}\stepcounter{equation}
\ee
in agreement with Ref.~\cite{Purcell46}. Note that using the spectral representation of the GF in the form of \Eq{GFalt}, the spontaneous decay rate $\gamma(\omega)$ of an inhomogeneous open optical system also scales like $\omega^3$ at $\omega\to 0$ [$I(\br_d,\omega)\propto \omega$ as demonstrated above]. This makes the Purcell factor (PF) $F(\omega)=\gamma(\omega)/\gamma_0(\omega)$ finite at $\omega\to 0$.

\section{Spectral representation of the Green's function and normalization of resonant states}
\label{SM-II}

The Green's function $\GF(\br,\br';\omega)$ of an open optical system is a tensor which satisfies Maxwell's wave equation with a delta-function source term (below $c=1$ is used for brevity of notations),
 \be
- \nabla\times\nabla\times \GF(\br,\br';\omega)+\omega^2\heps({\bf r;\omega})\GF(\br,\br';\omega)=\hat{
\mathbf{1}}\delta(\br-\br')\,,
\tag{S\theequation}\label{GFequ}\stepcounter{equation}
\ee and outgoing wave boundary conditions.
Treating $\GF(\br,\br';\omega)$ as a function of a complex $\omega$ we use the fact that the GF has a countable number of simple poles in the lower half plane at $\omega=\tilde{\omega}_n$. We further note that for $\omega\to \infty$, the GF vanishes inside the area of inhomogeneity of $\heps({\bf r};\omega)$. Note that the frequency dependence of the permittivity tensor is included. Then according to the  Mittag-Leffler theorem~\cite{More71,Bang81} we can write
\be
\GF(\br,\br';\omega)=\sum_n\frac{\Rn(\br,\br')}{\omega-\tilde{\omega}_n}\,,
\tag{S\theequation}\label{ML1}\stepcounter{equation}
\ee
where $\Rn(\br,\br')$ is the residue of the GF at $\omega=\tilde{\omega}_n$.

Now, for each RS having the eigenfrequency $\omega_n$ and the electric field $\En(\br)$ satisfying the homogeneous Maxwell wave equation
 \be
- \nabla\times\nabla\times \En(\br)+\omega_n^2\heps({\bf r};\omega_n)\En(\br)=0
\tag{S\theequation}\label{RSequ}\stepcounter{equation}
\ee
and outgoing wave boundary conditions, we introduce an analytic continuation $\Fn(\br,\omega)$,
such that
\be
\lim_{\omega\to\omega_n} \Fn(\br,\omega)=\En(\br)\,.
\tag{S\theequation}\label{lim}\stepcounter{equation}
\ee
$\Fn(\br,\omega)$ is defined as a solution of the inhomogeneous Maxwell wave equations
\be
- \nabla\times\nabla\times\Fn({\bf r};\omega)+\omega^2\heps(\br;\omega)\Fn(\br;\omega)=(\omega^2-\omega^2_n)\bs_n(\br)\,,
\tag{S\theequation}\label{Fequ}\stepcounter{equation}
\ee
in which $\bs_n(\br)$ is an arbitrary function vanishing outside the system and normalized in such a way that
\be
\int_{\cal V}\En(\br)\cdot\bs_n(\br)\,d\br=1\,,
\tag{S\theequation}\label{sigma}\stepcounter{equation}
\ee
where ${\cal V}$ is an arbitrary simply connected volume including all the inhomogeneities of $\heps({\bf r};\omega)$.
In the case of degenerate modes, $\omega_m=\omega_n$ for $m\neq n$, the source $\bs_n(\br)$ has to be chosen in such a way that, additionally,
$ \int_{\cal V}{\bf E}_m(\br)\cdot\bs_n(\br)\,d\br=0\,.$
Solving \Eq{Fequ} with the help of the GF \Eq{ML1} we obtain
\be
\Fn({\bf r};\omega)=\sum_{n'} \frac{\omega^2-\omega^2_n}{\omega-\tilde{\omega}_{n'}} \int_{\cal V} \Rnp(\br,\br')\bs_n(\br')\,d\br'\,.
\tag{S\theequation}\stepcounter{equation}
\ee
Taking the limit of \Eq{lim} and using the fact that $\GF(\br,\br';\omega)$ is a symmetric tensor, which follows from the reciprocity theorem~\cite{Born99}, we find
\be
\Rn(\br,\br')=\frac{\En(\br)\otimes\En(\br')}{2\omega_n}
\tag{S\theequation}\label{QnE}\stepcounter{equation}
\ee
and $\omega_n=\tilde{\omega}_n$, leading to the spectral representation \Eq{GFalt}. Substituting it into \Eq{GFequ} and using \Eq{RSequ}
results in the closure relation
\be
\sum_n\frac{\omega^2\heps(\br;\omega)-\omega_n^2\heps(\br;\omega_n)}{2\omega_n(\omega-\omega_n)}\,\En(\br)\otimes\En(\br')=
\hat{\mathbf 1}\delta(\br-\br')\,,
\tag{S\theequation}\label{Closure}\stepcounter{equation}
\ee
which in the absence of frequency dispersion of $\heps(\br)$ splits into
the sum rule \Eq{sumrule} and a simpler closure relation
\be
\frac{1}{2}\,\heps(\br)\sum_n\En(\br)\otimes\En(\br')=
\hat{\mathbf 1}\delta(\br-\br')\,.
\tag{S\theequation}\label{Closure2}\stepcounter{equation}
\ee
As already noted in Sec.~\ref{SM-I}, combining \Eq{GFalt} and the sum rule \Eq{sumrule} leads to an alternative form of the spectral representation \Eq{GFn} which was used in the resonant-state expansion (RSE)~\cite{MuljarovEPL10,DoostPRA14}.

The form of the GF \Eq{GFn} determines the normalization of RSs which technically follows from \Eq{sigma} by substituting $\bs_n(\br)$ from \Eq{Fequ} and taking the limit $\omega\to\omega_n$ (below the argument $\br$ is omitted for brevity):
\begin{align}
1=&\ \int_{\cal V}d\br\,\En\cdot\bs_n \nonumber\\
=&\ \lim_{\omega\to\omega_n} \int_{\cal V}d\br\, \En \cdot \frac{- \nabla\times\nabla\times\Fn+\omega^2\heps(\omega)\Fn}{\omega^2-\omega_n^2}\nonumber\\
&\ -\lim_{\omega\to\omega_n} \int_{\cal V}d\br\, \Fn \cdot \frac{- \nabla\times\nabla\times\En+\omega_n^2\heps(\omega_n)\En}{\omega^2-\omega_n^2}\nonumber\\
=&\ \lim_{\omega\to\omega_n} \int_{\cal V}d\br\, \Fn \cdot \frac{\omega^2\heps(\omega)-\omega_n^2\heps(\omega_n)}{\omega^2-\omega_n^2}\En\nonumber\\
&\ + \lim_{\omega\to\omega_n} \frac{\int_{\cal V}(\Fn\cdot\nabla\times\nabla\times \En -\En\cdot\nabla\times\nabla\times \Fn)d\br}{\omega^2-\omega_n^2}\nonumber\\
=&\ \int_{\cal V} d\br\, \En \cdot \left. \frac{\partial (\omega^2\heps(\omega))}{\partial (\omega^2)}\right|_{\omega=\omega_n}\!\!\En \nonumber\\
&\ + \lim_{\omega\to\omega_n}\frac{\displaystyle \oint _{S_{\cal V}} dS
\left(\En\cdot\frac{\partial\Fn}{\partial s}-\Fn\cdot\frac{\partial\En}{\partial
s}\right)}{\omega^2-\omega_n^2}\,,
\tag{S\theequation}\label{norm-der}\stepcounter{equation}
\end{align}
where after using some vector algebra we have applied the divergence theorem to convert a volume integral into a surface integral over the closed surface $S_{\cal V}$, the boundary of ${\cal V}$, with $\partial/\partial s$ denoting the directional derivative normal to this surface.

For any surface $S_{\cal V}$, the limit in the last term in \Eq{norm-der} can be evaluated explicitly by using the functional dependence of the electric field outside the system, where $\heps(\br)=\hat{\mathbf{1}}$ up to a scalar constant. For any mode with $\omega_n\neq 0$, the wave function of the RS
is given by $ \En(\br)={\bf Q}_n(\omega_n \br)$, where ${\bf Q}_n({\bf q})$ is a vector function satisfying the equation
\be
\nabla_{\bf q}\times\nabla_{\bf q}\times {\bf Q}_n({\bf q})={\bf Q}_n ({\bf q})
\tag{S\theequation}\label{F-equ}\stepcounter{equation}
\ee
and the proper boundary conditions at system interfaces and at ${\bf q}\to \infty$.
The analytic continuation of $\En(\br)$ can therefore be taken in the form \be
\Fn(\br,\omega)={\bf Q}_n(\omega\br)\,.
\tag{S\theequation}\label{F-Q}\stepcounter{equation}
\ee
In doing this, one could require, for example, that $\Fn(\br,\omega)$ outside the system has the same expansion
in terms of vector spherical harmonics as $\En(\br)$ itself, so that the frequency dependence of $\Fn(\br,\omega)$ comes only from the argument $\omega\br$ of the vector spherical harmonics, and not from the expansion coefficients.
This imposes certain condition on the the choice of $\sigma_n(\br)$ which was up to now treated as an arbitrary function.
Using the Taylor expansion of \Eq{F-Q} about the point $\omega=\omega_n$,
\begin{align}
\Fn(\br,\omega)\approx &\ {\bf Q}_n(\omega_n\br)+(\omega-\omega_n)\left.\frac{\partial {\bf Q}_n(\omega\br)}{\partial \omega}\right|_{\omega=\omega_n}
\nonumber\\
=&\ \En(\br)+\frac{\omega-\omega_n}{\omega_n} (\br\cdot\nabla)\,\En(\br)\,,
\tag{S\theequation}\label{ETaylor}\stepcounter{equation}
\end{align}
and substituting it into \Eq{norm-der} we obtain
\begin{align}
1=&\ \int_{\cal V}d\br\, \En \cdot \left. \frac{\partial (\omega^2\heps(\omega))}{\partial (\omega^2)}\right|_{\omega=\omega_n}\!\!\En \nonumber\\
&
+\frac{c^2}{2\omega^2_n}\oint_{S_{\cal V}} dS \left[\En\cdot\frac{\partial\Kn}{\partial s}-\Kn \cdot \frac{\partial \En}{\partial s}\right]
\tag{S\theequation}\label{norm-disp}\stepcounter{equation}
\end{align}
where
\be
\Kn(\br)=(\br\cdot\nabla)\,\En(\br)\,,
\tag{S\theequation}\label{K-equ}\stepcounter{equation}
\ee
and the speed of light is restored.

We note that the normalization of static modes \mbox{($\omega_n=0$)} is different and has been treated in Ref.~\cite{DoostPRA14}. They do not contribute to the radiative decay and thus are not further considered here.

In the absence of dispersion, the first integral in the normalization \Eq{norm-disp} is simplified to $\int_{\cal V} d\br\, \En \cdot \heps \En $, yielding  Eq.~(8) of the main text.
Using also spherical coordinates in the surface integral, \Eq{norm-disp} reduces to
\begin{align}
1=&\ \int_{\cal V}d\br\, \En \cdot \heps \En \tag{S\theequation}\label{norm-disp2}\stepcounter{equation}\\
&
+\frac{c^2}{2\omega^2_n}\oint_{S_{\cal V}} dS \left[\En\cdot\frac{\partial}{\partial s}r\frac{\partial\En}{\partial r}-r\frac{\partial \En}{\partial r} \cdot \frac{\partial \En}{\partial s}\right],
\nonumber
\end{align}
where $r=|\br|$ is the radius in spherical coordinates. Finally, if $\heps(\br)=\hat{\mathbf{1}} \varepsilon(\br)$ and the surface of integration is chosen in the form of a sphere of radius $R$ with the center at the origin, the normalization \Eq{norm-disp} takes its original form published in \Onlinecite{MuljarovEPL10}:
\begin{align}
1=&\ \int_{{\cal V}_R}  \varepsilon(\br)\En^2\, d\br \tag{S\theequation}\label{norm-disp3}\stepcounter{equation}\\
&
+\frac{c^2}{2\omega^2_n}\oint_{S_{{\cal V}_R}} \left[\En\cdot\frac{\partial}{\partial r}r\frac{\partial\En}{\partial r}-r\left(\frac{\partial \En}{\partial r}\right)^2\right] dS\,,
\nonumber
\end{align}
where ${\cal V}_R$ is the volume and $S_{{\cal V}_R}$ is the surface area of the sphere. Let us also note that if the homogeneous space outside the system is not vacuum but a medium described by a frequency-dependent uniform permittivity $\varepsilon(\omega)$, the surface term of the normalization  \Eq{norm-disp}
acquires an additional factor
\be
\left.\frac{1}{\varepsilon(\omega)}\frac{\partial ( \omega^2\varepsilon(\omega))}{\partial (\omega^2)}\right|_{\omega=\omega_n}
\tag{S\theequation}\stepcounter{equation}
\ee
which can be easily obtained following the derivation in \Eq{ETaylor}. Obviously this factor is equal to 1 for non-dispersive media.

The normalization of RSs with the surface term as given e.g. in \Eq{norm-disp} contains second-order spatial derivatives of the RS field. In some numerical implementations like those considered in Sec.\,\ref{Sec-Comsol} below, using second-order derivatives can lead to an accumulation of numerical errors. It is therefore useful to provide a version of normalization which contains only first derivatives of the field. This can be obtained by noting that the surface integral in \Eq{norm-disp} actually presents the flux through a closed surface $S_{\cal V}$ of the field
\be
{\bf \Phi}_1(\br) =\sum_{i=x,y,z} \left(E_i\nabla K_i -  K_i\nabla E_i\right)
\tag{S\theequation}\label{Phi}\stepcounter{equation}
\ee
in which $\bE=\En$ and ${\bf K}=\Kn$ defined by \Eq{K-equ}.
Then, using Gauss's theorem, this surface integral can then be transformed to the flux of another field ${\bf \Phi}_2$ linked to the first one through
\be
{\bf \Phi}_1={\bf \Phi}_2+\nabla \times {\bf A}\,,
\tag{S\theequation}\stepcounter{equation}
\ee
where ${\bf A}(\br)$ is an arbitrary field. By representing ${\bf \Phi}_1$ as
\be
{\bf \Phi}_1 = \nabla (\bE \cdot {\bf K})-2 \sum_{i=x,y,z}  K_i\nabla E_i\,,
\tag{S\theequation}\label{Phi1}\stepcounter{equation}
\ee
and then using the fact that
\begin{align}
2\nabla(\bE \cdot {\bf K})=&\nabla(\br\cdot\nabla) E^2
\tag{S\theequation}\label{EK}\stepcounter{equation}\\
=& -\nabla E^2 +\br \nabla^2 E^2 - \nabla\times(\br\times \nabla E^2)\,,
\nonumber
\end{align}
where $ E^2=\bE\cdot \bE$, second-order derivatives of the E-field are partly removed from \Eq{Phi1} by the curl. The remaining term containing $\nabla^2 E^2$  can be transformed to lower orders by applying Maxwell's wave equation in vacuum, $\nabla^2 \bE=-k^2 \bE$, so that
\be
\frac{1}{2}\nabla^2 E^2=-k^2 E^2 + \sum_{i,j=x,y,z} \left(\frac{\partial E_i}{\partial x_j}\right)^2.
\tag{S\theequation}\stepcounter{equation}
\ee
Then we obtain
\be
{\bf A}(\br)=-\frac{1}{2}\,\br\times \nabla E^2
\tag{S\theequation}\stepcounter{equation}
\ee
and
\be
{\bf \Phi}_2 =-\frac{1}{2}\nabla E^2-k^2 \br E^2 +\br\sum_{ij} \left(\frac{\partial E_i}{\partial x_j}\right)^2 -2\sum_{i}   K_i\nabla E_i
\tag{S\theequation}\label{Phi2}\stepcounter{equation}
\ee
($i,j=x,y,z$), the latter not containing derivatives of the field $\bE$ higher than first order. Omitting for simplicity of notation the dispersion in the volume integral, the modified normalization, equivalent to \Eq{norm-disp} but containing only first derivatives of $\En(\br)$ in the surface term, takes the form
\be
1=\ \int_{\cal V}d\En \cdot \heps \En\,d\br + \frac{c^2}{2\omega^2_n}\oint_{S_{{\cal V}}} {\bf \Phi}_2 \cdot d{\bf S} \,,
\tag{S\theequation}\label{norm-disp4}\stepcounter{equation}\\
\ee
where  ${\bf \Phi}_2(\br) $ is given by \Eq{Phi2}, with $\bE(\br)=\En(\br)$ and ${\bf K}(\br)=\Kn(\br)$ defined by \Eq{K-equ}.

\section{Leung-Kristensen normalization}
\label{Sec-Krist}

Following Leung {\it et al.}~\cite{LeungJOSAB96}, Kristensen {\it et al}.~\cite{KristensenOL12} have introduced a normalization of RSs in the form of Eq.\,(4) of the main text, which we call here Leung-Kristensen (LK) normalization. We found that this normalization is only correct for so-called $s$-waves, i.e. $l=0$ modes of a spherically symmetric system, where $l$ is the orbital quantum number. However, owing to the vectorial nature of the electromagnetic field, $l=0$ eigenmodes do not exist, so that the LK normalization is incorrect for all modes in electrodynamics.

We illustrate this finding for transverse-electric (TE) modes of a dielectric sphere. We compare the mode volume $V_n^{\rm LK}$, calculated using the LK normalization, with the correct one, $V_n$, calculated using the exact normalization Eq.~(8), by considering the relation
\be
V_n^{\rm LK}=V_n N_n(R)\,,
\tag{S\theequation}\label{Vratio}\stepcounter{equation}
\ee
in which $N_n(R)$ is the factor between the two volumes. It is given by the sum of the volume and surface normalization integrals in Eq.\,(4):
\be
N_n(R)= I_n(R)+S_n(R)
\tag{S\theequation}\stepcounter{equation}
\ee
with
\be
I_n(R)=\int_{{\cal V}_R} \varepsilon(\br) {\bf E}_n^2 (\br) d{\bf r}
\tag{S\theequation}\stepcounter{equation}
\ee
and
\be
S_n(R)= \frac{i}{2k_n} \oint_{S_R} {\bf E}_n^2 (\br) dS\,.
\tag{S\theequation}\stepcounter{equation}
\ee
Here ${\cal V}_R$ is the volume of a sphere of radius $R$, $S_R$ is its surface and $k_n=\omega_n/c$ is the RS wave number.
The LK normalization requires that the factor $N_n(R)$ is unity in the limit $R\to\infty$. To illustrate the resulting error, we calculate $N_n(R)$ for the correctly normalized $\En(\br)$ for finite $R$ and in the limit $R\to\infty$.

For a dielectric sphere of radius $a$ in vacuum, described by the dielectric constant
\be
\varepsilon(r)= \left\{
\begin{array}{cl}
n_r^2 & {\rm for}\ \  r\leqslant a \\
1 & {\rm for}\ \ r>a\,,\\
\end{array} \right.
\tag{S\theequation}\label{eps}\stepcounter{equation}
\ee
the eigenfunctions of the TE modes normalized via Eq.~(8)
have the form (in spherical polar coordinates)~\cite{DoostPRA14}:
\be
\mathbf{E}^{\rm TE}_n(\br)=A_l^{\rm TE}R_l(r,k_n)\begin{pmatrix}
0\\[5pt]
\dfrac{1}{\sin\theta}\dfrac{\partial}{\partial\varphi}Y_{lm}(\Omega)\\[10pt]
-\dfrac{\partial}{\partial\theta}Y_{lm}(\Omega)\\
\end{pmatrix},
\tag{S\theequation}\label{eqn:E_TE}\stepcounter{equation}
\ee
where $Y_{lm}(\Omega)$ are the spherical harmonics,
\be
R_l(r,k)= \left\{
\begin{array}{lll}
j_l(n_r k r)/j_l(n_r k a) & {\rm for} & r\leqslant a\, \\
h_l(kr)/h_l(ka) & {\rm for} & r>a\,, \\
\end{array}
\right.
\tag{S\theequation}\label{R-analyt}\stepcounter{equation}
\ee
$j_l(z)$ and $h_l(z)\equiv h_l^{(1)}(z)$ are, respectively, the spherical Bessel and Hankel functions of first kind,
\be
A^{\rm TE}_{l}=\sqrt{\frac{2}{l(l+1)a^3(n_r^2-1)}}
\tag{S\theequation}\label{A-normTE}\stepcounter{equation}
\ee
are normalization constants and $k_n$ are the solutions of the secular equation
\be
\frac{n_r j_{l+1}(n_r k_na)}{j_l(n_r k_na)}=\frac{h_{l+1}(k_na)}{h_l(k_na)}\,.
\tag{S\theequation}\label{secularTE}\stepcounter{equation}
\ee
Using these properties, we evaluate the volume and surface normalization integrals for $R\geqslant a$ as
\begin{align}
&I_n(R)=l(l+1)A_l^2\int_0^R R_l^2(r,k_n)\varepsilon(r) r^2 dr \nonumber\\
=&\frac{2}{a^3(n_r^2-1)}\left[n_r^2 \int_0^a\!\frac{j_l^2(n_r k_nr)}{j_l^2(n_r k_n a)} r^2 dr+\!\!\int_a^R\!\frac{h_l^2(k_nr)}{h_l^2(k_n a)} r^2 dr\right]\nonumber\\
=&\ 1+\frac{(R/a)^3}{n_r^2-1}\frac{h_l^2(k_nR)}{h_l^2(k_n a)}\left[1-\frac{h_{l-1}(k_nR)h_{l+1}(k_nR)}{h_l^2(k_n R)}\right]
\tag{S\theequation}\stepcounter{equation}
\end{align}
and
\be
S_n(R)=\frac{iR^2}{2k_n}l(l+1)A_l^2 R_l^2(R,k_n) =\!\frac{i}{k_n R}\frac{(R/a)^3}{n_r^2-1}\frac{h_l^2(k_nR)}{h_l^2(k_n a)}\,.
\tag{S\theequation}\stepcounter{equation}
\ee
and consequently find
\be
N_n(R)= 1+\frac{1}{n_r^2-1}\left(\frac{R}{a}\right)^3\frac{h_l^2(k_nR)}{h_l^2(k_n a)} Q_n(R)\,,
\tag{S\theequation}\stepcounter{equation}
\ee
where
\be
 Q_n(R)=1-\frac{h_{l-1}(k_nR)h_{l+1}(k_nR)}{h_l^2(k_n R)}+\frac{i}{k_n R}\,.
\tag{S\theequation}\stepcounter{equation}
\ee
To investigate the behaviour of $N_n(R)$ for large $k_nR$, we use the asymptotic formula for $h_l(z)$ at large arguments. We find that in $Q_n(R)$, the 0th-, 1st-, and 2nd-order terms in $1/(k_nR)$ are vanishing, so that
\be
Q_n(R)= \frac{C_l(k_nR)}{(k_nR)^3}
\tag{S\theequation}\stepcounter{equation}
\ee
and consequently
\be
\frac{V_n^{\rm LK}}{V_n}=N_n(R)=1+\frac{C_l(k_nR)}{(n_r^2-1)(k_na)^3 h_l^2(k_n a)}\,\frac{e^{2ik_nR}}{(k_n R)^2}\,,
\tag{S\theequation}\label{Vratio1}\stepcounter{equation}
\ee
where $C_l(z)=-i l(l+1)/2 +{\cal O}(1/z)$. Similarly, for the normalization without surface term, i.e. with the volume term only, Eq.\,(3) of the main text, we find
\be
\frac{V_n^{\rm vol}}{V_n}=I_n(R)=1+\frac{D_l(k_nR)}{(n_r^2-1)(k_na)^3 h_l^2(k_n a)}\,e^{2ik_nR}\,,
\tag{S\theequation}\label{Vratio2}\stepcounter{equation}
\ee
where $D_l(z)=(-1)^{l}i +{\cal O}(1/z^2 )$. Clearly, Eq.\,(4) brings an improvement compared to Eq.\,(3) -- the last term in \Eq{Vratio1} is decreasing with $R$ for Q $\gg 1$ modes, such as whispering gallery modes (WGMs), for which $|e^{2ik_n(R-a)}|\approx 1$ up to rather large $R$. However both normalizations diverge for $R\to\infty$ due to the exponential factor $e^{2ik_n R}$.

\begin{figure}[t]
\includegraphics*[width=\columnwidth]{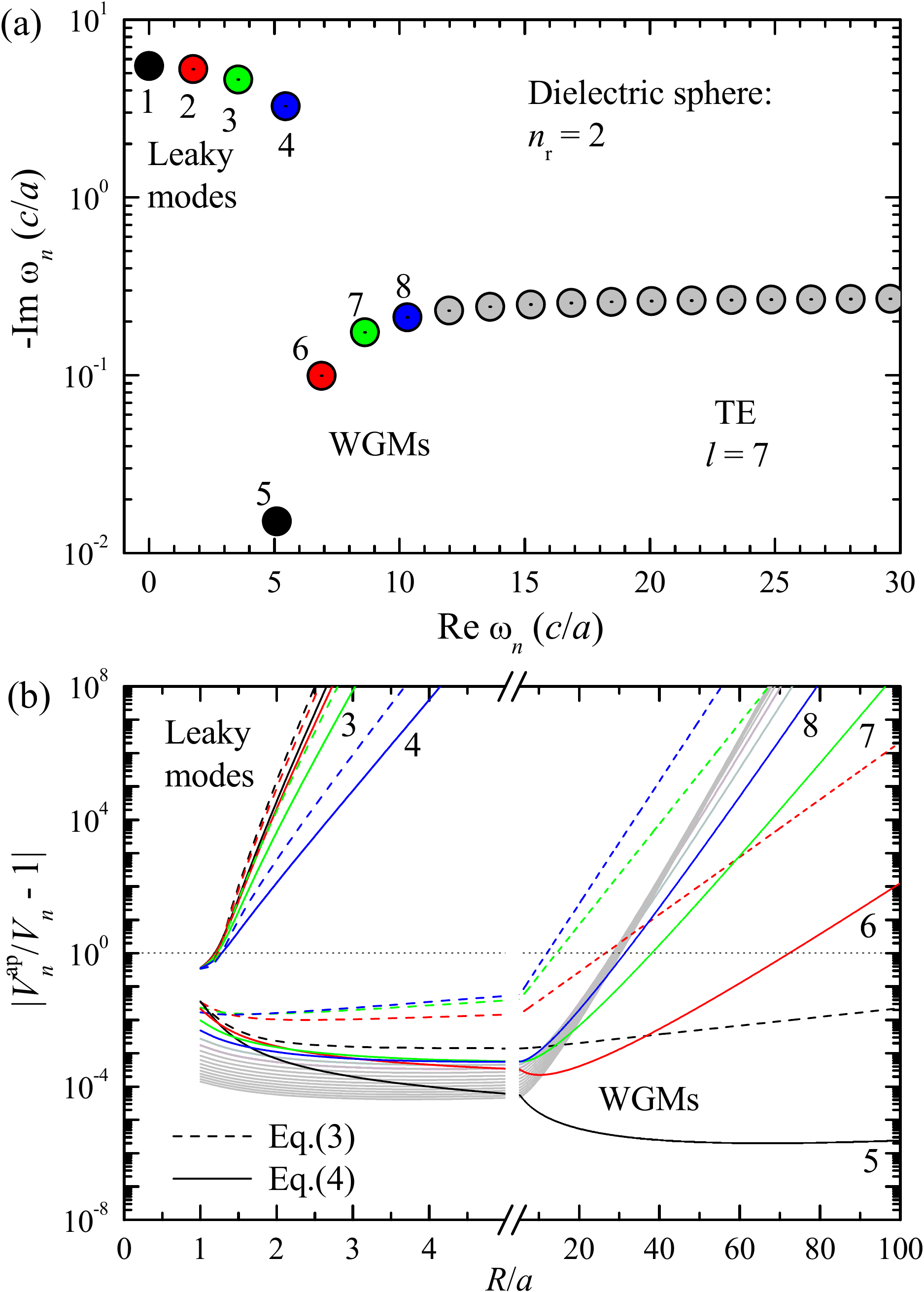}
\caption{(a) Frequencies of $l = 7$ TE modes (RSs) of a dielectric sphere in vacuum,
with permittivity $\epsilon = 4$ and radius $a$. (b) Relative error of the approximate mode volume $|V_n^{\rm ap}/V_n-1|$ as function of the radius $R$ of the sphere of integration, for the modes shown in (a). $V_n^{\rm ap}$ is calculated using Eq.~(3), having no surface term, and Eq.~(4), having the incorrect surface term used in the literature \cite{LeungJOSAB96,KristensenOL12}.
}\label{fig:SM1}
\end{figure}
\begin{figure}[t]
\includegraphics*[width=\columnwidth]{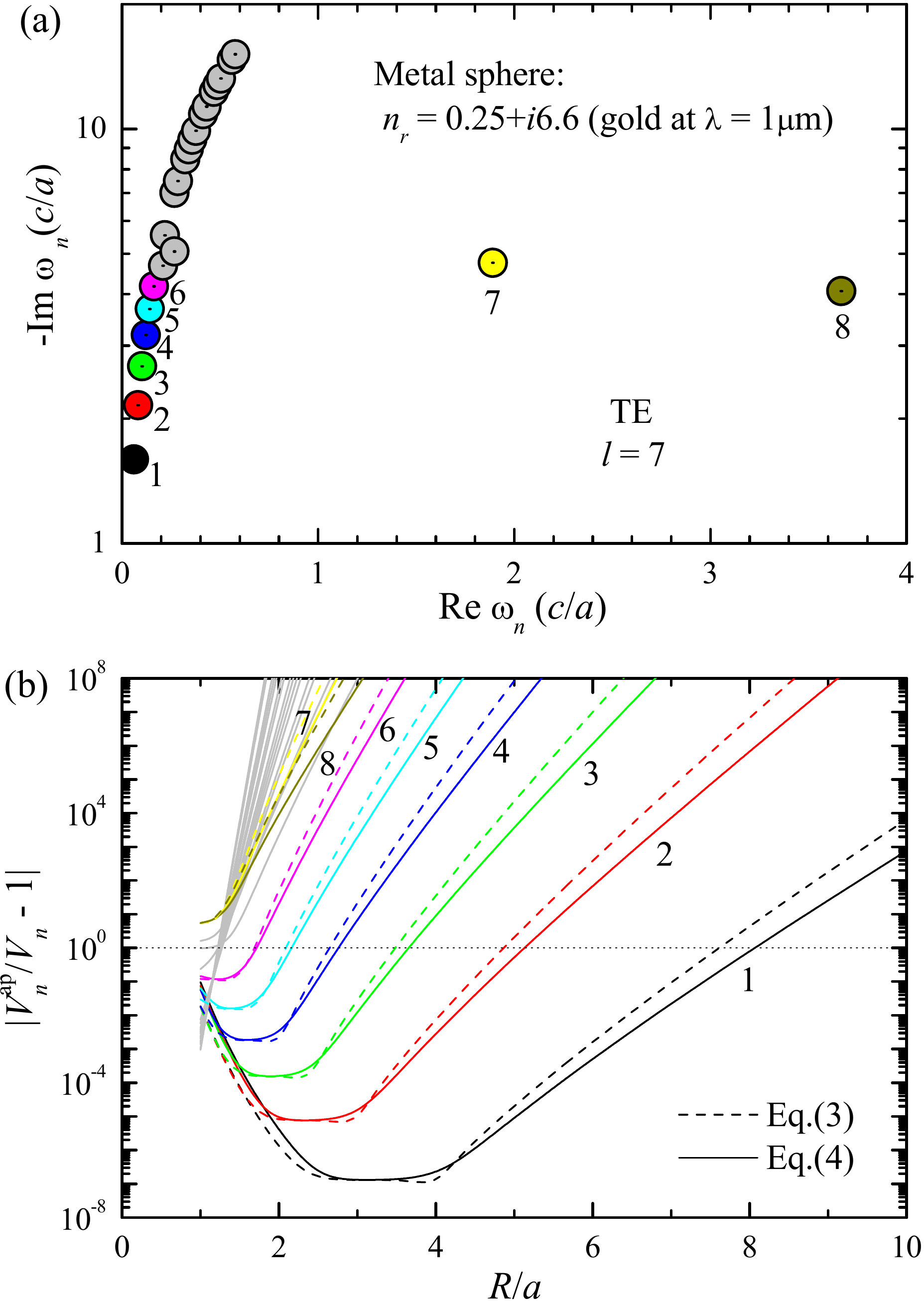}
\caption{As Fig.~\ref{fig:SM1}, but for a sphere with a fixed complex dielectric constant $\epsilon=-43.5+i3.33$, equal to that of gold at a light wavelength of $\lambda=1\,\mu$m.
}\label{fig:SM2}
\end{figure}
This is exemplified in Figs.~\ref{fig:SM1}(b) and \ref{fig:SM2}(b) where the relative errors in the mode volume, $V_n^{\rm ap}/V_n-1$ (with $V_n^{\rm ap}=V_n^{\rm LK}$ or $V_n^{\rm ap}=V_n^{\rm vol}$), are shown for several RSs of a dielectric and a metal sphere, respectively, with corresponding eigenfrequencies given in Figs.~\ref{fig:SM1}(a) and \ref{fig:SM2}(a). In Fig.~\ref{fig:SM1}(b), the strongest deviation and exponentially growing errors are seen for leaky modes already for small values of $R$. For WGMs the errors can be small up to rather large $R$, showing an apparent convergence, in agreement with the analytic treatment given above, and the advantage of using Eq.\,(4) versus Eq.\,(3) is clearly observed. Nevertheless, the error diverges also for WGMs in the limit $R\to\infty$, in agreement with the asymptotics given by Eqs.\,(\ref{Vratio1}) and (\ref{Vratio2}). Moving to the metal sphere we observe that the modes have typically a low $Q$, such that the exponential divergence of the error with $R$ is more pronounced. For some low-frequency modes (labeled 1-5) the error initially decays exponentially up to a finite $R$ where the error is minimized. We note that this is observed both for Eq.\,(4) and Eq.\,(3), indicating that the surface term is not the relevant aspect here, and we find that it actually increases the error at small $R$. These states are quasi-bound states in the metal sphere which are evanescent close to the sphere due to the angular momentum, similar to WGMs. At $R\gtrsim lc/|\omega_n|$, they become propagating, and the error recovers the expected exponential divergence.

The correct normalization Eq.\,(8) can be analyzed in a similar way. It consists of two terms, $I_n(R)$ and $J_n(R)$, where
\begin{align}
J_n(R)=&\frac{1}{2k_n^2}\oint_{S_R} dS \left[\En\cdot\frac{\partial}{\partial r}r\frac{\partial\En}{\partial r}-r\left(\frac{\partial \En}{\partial r}\right)^2\right]\nonumber\\
=&\frac{1}{2k_n^2}\int d\Omega \left[\frac{1}{\sin^2\theta}\left(\frac{\partial Y_{lm}}{\partial\varphi}\right)^2+\left(\frac{\partial Y_{lm}}{\partial\varphi}\right)^2\right]\nonumber\\
&\times A_l^2 R^2\left[ R_l\frac{\partial}{\partial r}r\frac{\partial R_l}{\partial r}-r\left(\frac{\partial R_l}{\partial r}\right)^2\right]_{r=R}
\nonumber
\end{align}
\be
=\frac{R^3}{a^3} \frac{zh_l(z)h_l'(z)+z^2h_l(z)h_l''(z)-z^2{h_l'}^2(z)}{(n_r^2-1)h_l^2(k_na) z^2}
\tag{S\theequation}\stepcounter{equation}
\ee
with $z=k_nR$. We thus obtain
\be
I_n(R)+J_n(R)=1+ \frac{R^3}{a^3(n_r^2-1)} \frac{p_l(z)}{h_l^2(k_na) z^2}\,,
\tag{S\theequation}\stepcounter{equation}
\ee
where
\begin{align}
p_l(z)=&\ zh_l(z)h_l'(z)+z^2h_l(z)h_l''(z)-z^2{h_l'}^2(z)
\nonumber\\
&+z^2 h_l^2(z)-z^2 h_{l-1}(z)h_{l+1}(z)=0\,,
\tag{S\theequation}\stepcounter{equation}
\end{align}
according to Bessel's equation and recursive relations for Hankel functions \cite{AbramowitzStegun} following from it. This confirms that  Eq.\,(8) provides the exact normalization condition $I_n(R)+J_n(R)=1$, independent of $R$.

\section{Comparison with Sauvan et al.}

In the normalization suggested by Sauvan {\it et al.}~\cite{SauvanPRL13} the electric field ${\bf E}$ of a RS (we drop here the index $n$) is normalized in such a way that
\be
I_1+I_2=1\,,
\label{norm-Sauvan}
\tag{S\theequation}\stepcounter{equation}
\ee
where
\be
2I_1=\int_{{\cal V}_1}d\br\, {\bf E} \cdot \frac{\partial (\omega \heps(\omega))}{\partial \omega}{\bf E}
-\int_{{\cal V}_1}d\br\, {\bf H} \cdot \frac{\partial (\omega\hmu(\omega))}{\partial \omega} {\bf H}
\label{I1def}
\tag{S\theequation}\stepcounter{equation}
\ee
is an integral over a volume ${\cal V}_1$ including the system inhomogeneity and $I_2$ is an integral of the same function over the region inside the perfectly matched layer (PML) in which the field decays due to the artificially absorbing medium of the PML.
Here ${\bf H}$ is the corresponding magnetic field of the RS. Note that we have used and extra factor of 2 in \Eq{I1def}, as compared to a similar equation in \Onlinecite{SauvanPRL13}). This is done in order make this normalization, which add up contributions of the electric and magnetic field,  comparable with our normalization \Eq{norm-disp}, dealing with the electric field only.  We compare our normalization \Eq{norm-disp} with \Eq{norm-Sauvan} by evaluating $I_1$, for which numerical values are provided in Ref.~\cite{SauvanPRL13} for a TM mode of a gold sphere with radius $a=0.1\,\mu$m having the wavelength $\lambda =2\pi c/\omega=(0.607+0.239i)\,\mu$m. We use the dielectric constant of gold in the Drude model with the same parameters as in Ref.~\cite{SauvanPRL13,Sauvan-private}:
$\epsilon(\omega)=1-\lambda^2/(0.15^2(1+i0.075\lambda))$ with $\lambda$ measured in $\mu$m.

The electric field of a TM mode has the form~\cite{DoostPRA14}
\be
\mathbf{E}(\br)=\dfrac{{A}^{\rm TM}_l(k)}{\varepsilon(r) k r}\left(
\begin{array}{ccc}
l(l+1)R_l(r,k)Y_{lm}(\Omega)\\[5pt]
\dfrac{\partial}{\partial r} r R_l(r,k)\dfrac{\partial}{\partial\theta}Y_{lm}(\Omega)\\[10pt]
\dfrac{\partial}{\partial r} \dfrac{r R_l(r,k)}{\sin\theta}\dfrac{\partial}{\partial\varphi}Y_{lm}(\Omega)\\
\end{array}
\right) \label{eqn:E_TM}
\tag{S\theequation}\stepcounter{equation}
\ee
in which $k=\omega/c$ is a solution of the secular equation for TM modes
\be
\frac{1}{n_r}\frac{j_{l+1}(n_r k a)}{j_l(n_r k a)}=\frac{h_{l+1}(k a)}{h_l(k a)}+\frac{l+1}{ka}\left(1-\frac{1}{n_r^2}\right),
\tag{S\theequation}\label{secularTM}\stepcounter{equation}
\ee
$R_l(r,k)$ is given by \Eq{R-analyt} and $\varepsilon(r)$ by \Eq{eps} with  $n_r^2=\epsilon(\omega)$, taking any of the two roots for $n_r$.
The normalization constant $A^{\rm TM}_l(k)$ calculated using the correct normalization \Eq{norm-disp} has the form
\be
\frac{n_r A^{\rm TE}_l}{A^{\rm TM}_l(k)\!\!}=\sqrt{\left[\frac{j_{l-1}(n_r ka)}{j_l(n_r ka)}-\frac{l}{n_r ka}\right]^2\!\!\!\!+\!\frac{l(l\!+\!1)}{k^2a^2}\!+\!\eta C_l(k)},
\label{A-normTM}
\tag{S\theequation}\stepcounter{equation}
\ee
where $A^{\rm TE}_l$ is given by \Eq{A-normTE}, and the last term under the square root takes into account the effect of the dispersion, with
\be
\eta=\frac{1}{\epsilon(\omega)}\frac{\partial(\omega^2 \epsilon(\omega))}{\partial(\omega^2)}-1
\tag{S\theequation}\stepcounter{equation}
\ee
and
\be
(n_r^2-1)C_l(k)=\frac{2(l\!+\!1)}{k^2a^2}+n_r^2\!\left[\frac{j_{l+1}^2(n_r ka)}{j_l^2(n_r ka)}-\frac{j_{l+2}(n_r ka)}{j_l(n_r ka)}\right].
\tag{S\theequation}\stepcounter{equation}
\ee
Note that the normalization constant $A^{\rm TM}_l(k)$ of a TM mode, defined by \Eq{A-normTM}, generalizes the one used for a dielectric sphere, which is given by Eq.~(29) of Ref.~\cite{DoostPRA14}, as it takes into account the dispersion of the metal via the term $\eta C_l(k)$.

The corresponding magnetic field of the same RS has the form
\be
i{\bf H}(\br)={A}^{\rm TM}_l(k) R_l(r,k)\left(
\begin{array}{ccc}
0\\[5pt]
\dfrac{1}{\sin\theta} \dfrac{\partial}{\partial \varphi}Y_{lm}(\Omega)\\[10pt]
-\dfrac{\partial}{\partial\theta}Y_{lm}(\Omega)\\
\end{array}
\right) .
\label{eqn:H_TM}
\tag{S\theequation}\stepcounter{equation}
\ee
The integral $I_1=(I_{1E}+I_{1H})/2$ over a sphere of radius $R\geqslant a$ is then evaluated in the following way:
\begin{align}
I_{1E}=& \int_{{\cal V}_R}d\br\, {\bf E} \cdot \frac{\partial (\omega \heps(\omega))}{\partial \omega}{\bf E}
\nonumber\\
=&\dfrac{\left[A^{\rm TM}_l(k)\right]^2\!\!\!}{k^2}\left\{ [l(l\!+\!1)]^2\!\!\!\int_0^R\!\!\! dr\frac{R_l^2(r,k) \beta(r)}{\varepsilon^2(r)}\!\!\int \!\!d\Omega Y^2_{lm}(\Omega)\right.
\nonumber\\
&+\int_0^R dr\frac{\beta(r)}{\varepsilon^2(r)}\left(\frac{\partial}{\partial r} r R_l(r,k)\right)^2
\nonumber\\
&\times\left.\int d\Omega\left[\left(\frac{\partial Y_{lm}(\Omega)}{\partial \theta}\right)^2+\frac{1}{\sin^2\theta}\left(\frac{\partial Y_{lm}(\Omega)}{\partial \varphi}\right)^2\right]\right\}
\nonumber\\
=&\dfrac{\left[A^{\rm TM}_l(k)\right]^2\!\!\!}{k^2} l(l\!+\!1)\!\!\! \int_0^R\!\!\!dr \frac{\beta(r)}{\varepsilon^2(r)} \left[ l(l\!+\!1) R_l^2+\left(\partial_r r R_l\right)^2\right]\!,
\label{I1E}
\tag{S\theequation}\stepcounter{equation}
\end{align}
where
\be
\beta(r)= \left\{
\begin{array}{cl}
\dfrac{\partial (\omega \epsilon(\omega))}{\partial \omega} & {\rm for}\ \  r\leqslant a \\[10pt]
1 & {\rm for}\ \ r>a\,,\\
\end{array} \right.
\label{beta}
\tag{S\theequation}\stepcounter{equation}
\ee
and
\begin{align}
I_{1H}=& -\int_{{\cal V}_R}d\br\, {\bf H} \cdot \frac{\partial (\omega \hmu(\omega))}{\partial \omega}{\bf H}
\nonumber\\
=&\left[A^{\rm TM}_l(k)\right]^2 \!\!\!\int_0^R\!\!\! r^2  R_l^2 dr \!\!\int \!d\Omega \left[\left(\partial_\theta Y_{lm}\right)^2
+\frac{(\partial_\varphi Y_{lm})^2}{\sin^2\theta}\right]
\nonumber\\
=&\left[A^{\rm TM}_l(k)\right]^2 l(l+1) \int_0^R r^2 R_l^2 dr \,,
\label{I1H}
\tag{S\theequation}\stepcounter{equation}
\end{align}
using $\hmu=\hat{\mathbf{1}}$ everywhere.
The integral in \Eq{I1E} is calculated analytically using the Bessel equation and integration by parts:
\begin{align}
\int dr &\left[ l(l +1)R_l^2+\left(\partial_r r R_l\right)^2\right] \nonumber\\
&= rR_l^2+r^2R_l\partial_r R_l+\varepsilon k^2\int r^2 R_l^2 dr\,,
\label{integral}
\tag{S\theequation}\stepcounter{equation}
\end{align}
where $\varepsilon$ is constant in each area of space. The integral in the last term of Eqs.~(\ref{I1H}) and (\ref{integral}) is a known analytic integral:
\be
\int x^2 f_l^2(\alpha x)dx=\frac{x^3}{2}\left[f_l^2(\alpha x)-f_{l-1}(\alpha x)f_{l+1}(\alpha x)\right]\,,
\tag{S\theequation}\stepcounter{equation}
\ee
in which $f_l(z)$ is any spherical Bessel function, $j_l(z)$ or $h_l(z)$, and $\alpha$ is a complex constant. We therefore find
\begin{align}
&I_{1E}=\dfrac{\left[A^{\rm TM}_l(k)\right]^2}{k^2}l(l+1) \left(
\frac{1}{\epsilon(\omega)}\frac{\partial(\omega \epsilon(\omega))}{\partial\omega} I_{1E}^a+I_{1E}^R\right),\nonumber\\
&I_{1H}=\dfrac{\left[A^{\rm TM}_l(k)\right]^2}{k^2}l(l+1) \left( I_{1H}^a+I_{1H}^R\right),
\label{bothI}
\tag{S\theequation}\stepcounter{equation}
\end{align}
where
\begin{align}
I_{1E}^a=&\ a(l+1)-\frac{n_r ka^2}{2}(2l+3)\frac{j_{l+1}(n_r k a)}{j_l(n_r k a)}
\nonumber\\
&\ +\frac{n^2_r k^2a^3}{2}\left(1+\frac{j^2_{l+1}(n_r k a)}{j^2_l(n_r k a)}\right)\,,
\nonumber\\
I_{1E}^R=& \frac{1}{2h^2_l(k a)}\left[-kr^2(2l+3)h_{l+1}(k r)h_{l}(k r)\right.
\nonumber\\
&\left. +2r(l+1)h^2_l(k r)+k^2r^3\left(h^2_l(k r)+h^2_{l+1}(k r)\right)
\right]_a^R,
\nonumber\\
I_{1H}^a=&\left[\frac{k^2r^3}{2j^2_l(n_r k a)}\left(j^2_l(n_r k r)\!-\!j_{l-1}(n_r k r)j_{l+1}(n_r k r)\right)\right]_0^a \!\!,
\nonumber\\
I_{1H}^R=&\left[\frac{k^2r^3}{2h^2_l(k a)}\left(h^2_l(k r)-h_{l-1}(k r)h_{l+1}(k r)\right)\right]_a^R\!.
\label{fourI}
\tag{S\theequation}\stepcounter{equation}
\end{align}

\begin{table}
\caption{The values of the integral \Eq{I1def} calculated in the present work ($I_1$) and the relative difference between $I_1$ and the value $I^{\rm S}_1$ calculated by Sauvan {\it et al.} \cite{SauvanPRL13}, for three integration radii $R$, for the  mode with the wavelength $2\pi c/\omega=(0.607+0.239i)\,\mu$m in a gold nanosphere of radius $a=0.1\,\mu$m. }
\begin{tabular}{|c|c|c|c|}
  \hline
  $R$ [$\mu$m] & $I_1$ & $I^{\rm S}_1/I_1-1$ & $|I^{\rm S}_1/I_1-1|$\\
  \hline
  0.15 & $0.61936187690$ & $(5.66-1.29\,i)$ & $5.80\times10^{-9}$\\
  &$-0.44899671324\,i$ &$\times 10^{-9}$ & \\
  1.0 & $6.56641919859$ & $(0.057+2.095\,i)$ & $2.095\times10^{-8}$\\
  & $+0.49127433385\,i$ & $\times 10^{-8}$& \\
  2.0 & $1052.29778832465$ & $(0.100+4.468\,i)$ & $4.469\times10^{-8}$\\
    &  $-1235.22683098918\,i$ & $\times10^{-8}$ & \\

\hline
\end{tabular}
\end{table}

The values of $I_1=(I_{1E}+I_{1H})/2$ calculated using Eqs.~(\ref{bothI}) and (\ref{fourI}) are shown in Table~S1 and compared with the values $I_1^{\rm S}$  provided by Sauvan {\it et al.}~\cite{SauvanPRL13,Sauvan-private} for the same radii $R$ of the sphere of integration. One can see an excellent agreement between the two approaches, with a relative error in the $10^{-7}-10^{-8}$ range. We note however that this result was obtained for a spherically symmetric system which is effectively one-dimensional (1D), for which the calculation in Ref.~\cite{SauvanPRL13} was done analytically~\cite{Sauvan-private} that actually explains the excellent agreement with the strict result. In a full 3D calculation the use of a PML may lead to more significant errors. For instance, using the approach of \Onlinecite{SauvanPRL13} a deviation of about 2\% of the PF from the direct numerical evaluation of the GF was found~\cite{SauvanPRL13,BaiOE13} for an optical mode in a gold rod with cylindrical symmetry. A more detailed comparison of this numerical normalization method with the exact normalization would be interesting.

\section{Details of the Purcell factor calculation}

In this section we provide some details of our calculation of the PF for a dielectric spherical resonator in vacuum, of radius $a$ and refractive index $n_r$; numerical results are presented here and in the main text. The PF is expressed in terms of the mode volumes via Eq.~(9), and the mode volume of a RS is given by Eq.~(2) of the main text, in terms of its normalized electric field. The latter has an explicit analytic form for a spherical resonator, which is given  by \Eq{eqn:E_TE} for TE and by \Eq{eqn:E_TM} for TM polarization. The normalization constants are given by Eqs.~(\ref{A-normTE}) and (\ref{A-normTM}). Static modes do not contribute to the PF as noted in Sec.~\ref{SM-I} and thus are not considered here.

\begin{figure}[t]
\includegraphics*[width=\columnwidth]{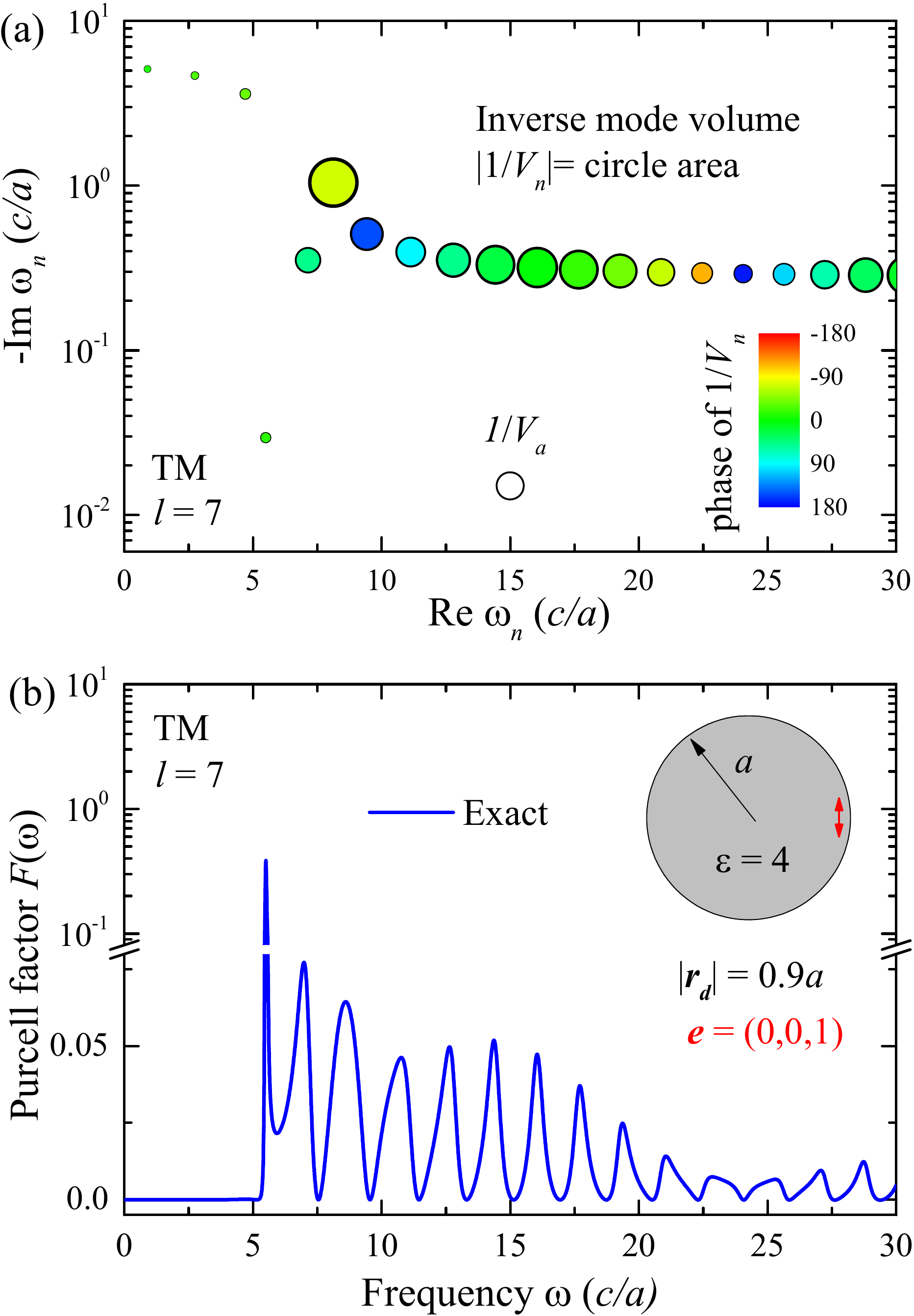}
\caption{(a) Mode volumes for a dielectric sphere in vacuum, with permittivity $\varepsilon =4$ and radius $a$, for $l=7$ TM modes, for a point dipole placed at $|\br_d|=0.9a$ with direction ${\bf e}=(0,0,1)$, in spherical coordinates (see sketch). The mode volume
is presented as the sum of the inverse mode volume over all degenerate states $m =-l,\dots,l$. Its amplitude is shown by
the circle area and its phase by the color. The volume of the sphere $V_a=4\pi a^3/3$ is shown for comparison. The position
of the circles in the complex frequency plane is given by the mode eigenfrequency $\omega_n$.
(b) Partial Purcell factor as a function of the dipole frequency $\omega$, calculated for the geometry of (a).
}\label{fig:SM3}
\end{figure}

\begin{figure}[t!]
\includegraphics*[width=\columnwidth]{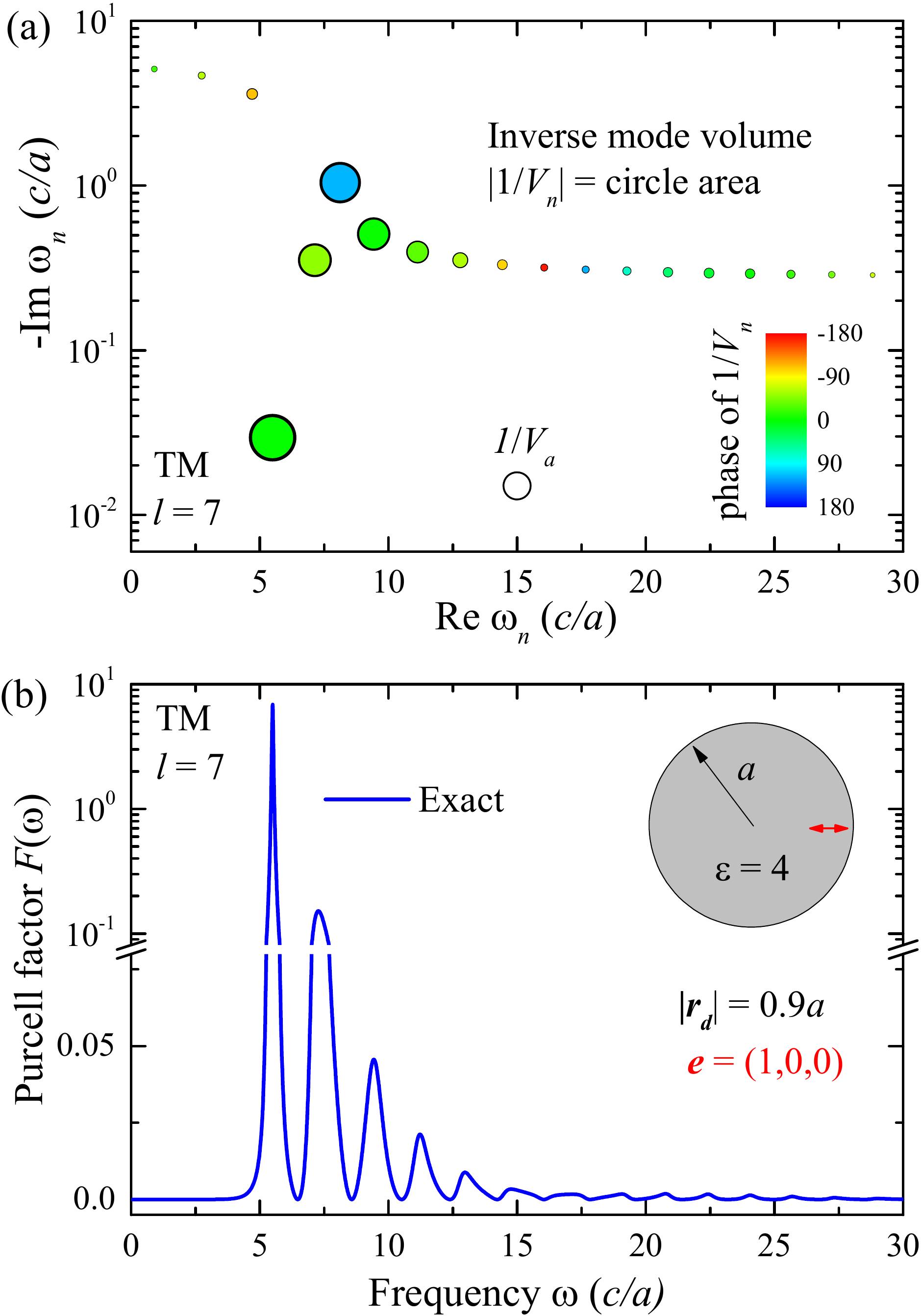}
\caption{As in Fig.\,\ref{fig:SM3} but for an orthogonal direction of the dipole: ${\bf e}=(1,0,0)$.
}\label{fig:SM4}
\end{figure}
Owing to the spherical symmetry of the resonator, RS eigenfrequencies are $2l+1$ degenerate with respect to the azimuthal quantum number  $m$ (here $l$ is the orbital quantum number). Therefore, for each set of degenerate RSs, we introduce a collective mode volume $V_l$ defined as
\be
\frac{\mu^2}{V_l}=\sum_{m=-l}^l \left[\bmu\cdot\bE_{lm}(\br)\right]^2\,,
\tag{S\theequation}\stepcounter{equation}
\ee
where the quantum numbers $l$ and $m$ are shown explicitly but the RS index $n$ is dropped for brevity of notation. Then, using the vector components of the dipole moment in spherical coordinates,
\be
\bmu=\mu_r{\bf e}_r+\mu_\theta{\bf e}_\theta+\mu_\varphi{\bf e}_\varphi\,,
\tag{S\theequation}\stepcounter{equation}
\ee
and the sum rules for spherical harmonics, we obtain
\begin{align}
\frac{\mu^2}{V_l^{\rm TE}}&= \dfrac{\left[E_l^{\rm TE}(r)\right]^2}{l(l+1)}\!\!\sum_{m=-l}^l\!
\left[\mu_\theta^2\left(\!\frac{1}{\sin\theta}\frac{\partial Y_{lm}}{\partial \varphi}\!\right)^{\!\!2}
\!\!+\!\mu^2_\varphi\!\left(\!\frac{\partial Y_{lm}}{\partial \theta}\!\right)^{\!\!2}
\right]
\nonumber\\
&=\left[E_l^{\rm TE}(r)\right]^2 \frac{2l+1}{8\pi} (\mu_\theta^2+\mu_\varphi^2)
\label{V-TE}
\tag{S\theequation}\stepcounter{equation}
\end{align}
for TE modes, and
\begin{align}
\frac{\mu^2}{V_l^{\rm TM}}=& \left[E_l^{\rm TM1}(r)\right]^2 l(l+1) \sum_{m=-l}^l\!
\mu_r^2 Y_{lm}^{2} \nonumber\\
&\!\!+\!\dfrac{\left[E_l^{\rm TM2}(r)\right]^2}{l(l+1)}\!\!\!\sum_{m=-l}^l\!\!
\left[\mu_\theta^2\!\left(\!\frac{\partial Y_{lm}}{\partial \theta}\!\right)^{\!\!2}
\!\!\!+\!\mu^2_\varphi\!\left(\!\frac{1}{\sin\theta}\frac{\partial Y_{lm}}{\partial \varphi}\!\right)^{\!\!2}
\right]
\nonumber\\
=&\left[E_l^{\rm TM1}(r)\right]^2 l(l+1) \frac{2l+1}{4\pi} \mu_r^2
\nonumber\\
&+\left[E_l^{\rm TM2}(r)\right]^2 \frac{2l+1}{8\pi} (\mu_\theta^2+\mu_\varphi^2)
\label{V-TM}
\tag{S\theequation}\stepcounter{equation}
\end{align}
for TM modes, where
\begin{align}
E_l^{\rm TE}(r)=&\sqrt{\frac{2}{(n_r^2-1)a^3}} R_l(r)\,,\nonumber\\
E_l^{\rm TM1}(r)=&\sqrt{\frac{2}{(n_r^2-1)a^3 D_l}} \frac{1}{n_r k r} R_l(r)\,,
\label{E-TETM}
\tag{S\theequation}\stepcounter{equation}\\
E_l^{\rm TM2}(r)=&\sqrt{\frac{2}{(n_r^2-1)a^3 D_l}} \frac{1}{n_r k r} \frac{\partial}{\partial r}r R_l(r)\,,
\nonumber
\end{align}
$R_l(r)= j_l(n_rk r)/j_l(n_r ka)$, $r$ is the position of the dipole (inside the dielectric sphere), and
\be
D_l=\left[\frac{j_{l-1}(n_r ka)}{j_l(n_r ka)}-\frac{l}{n_r ka}\right]^2+\frac{l(l+1)}{k^2a^2}\,.
\tag{S\theequation}\stepcounter{equation}
\ee
\begin{figure}[t!]
\includegraphics*[width=\columnwidth]{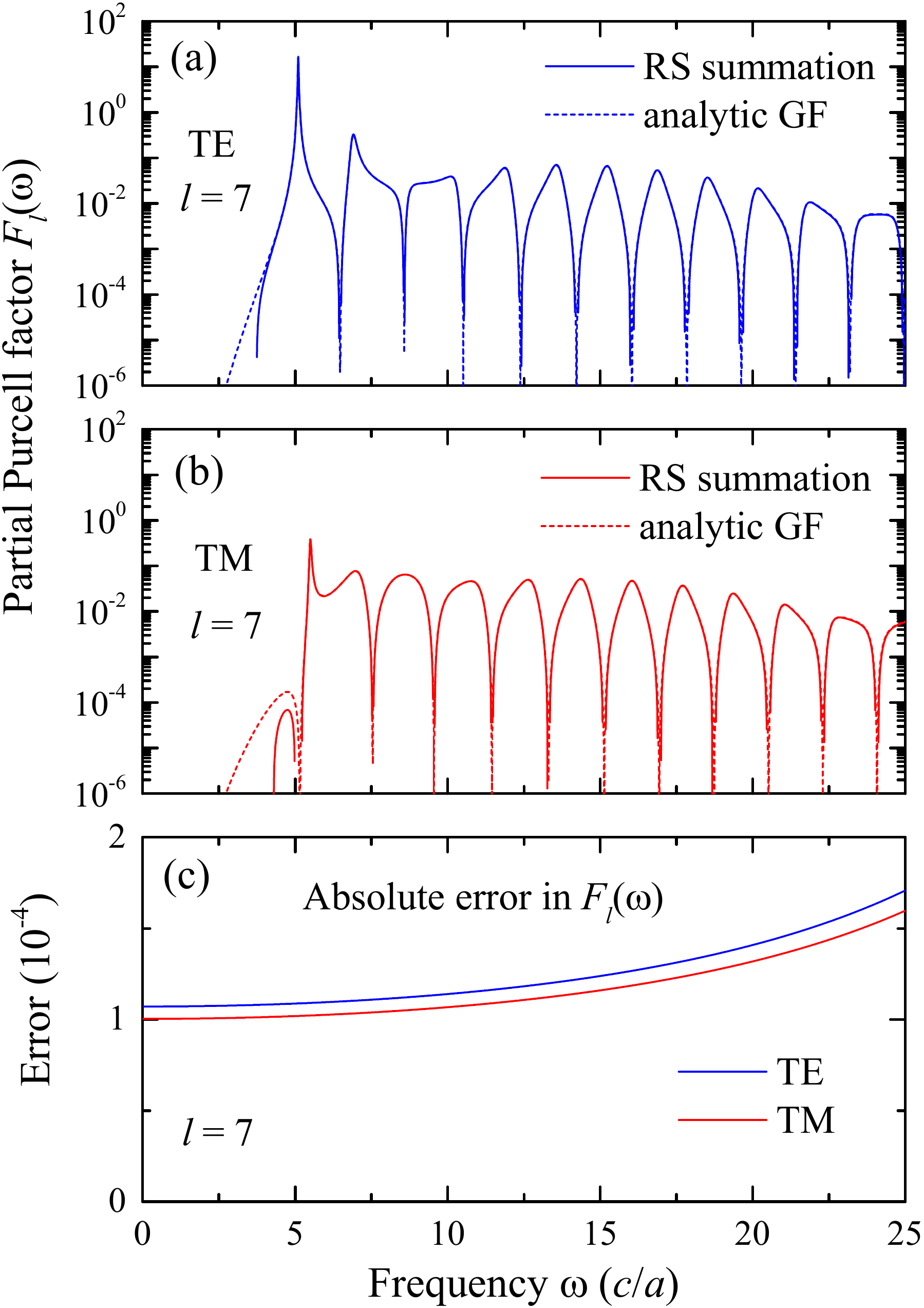}
\caption{(a) Partial Purcell factor of a dielectric sphere in vacuum, with permittivity $\varepsilon =4$ and radius $a$, for a point dipole placed at $|\br_d|=0.9a$ with direction ${\bf e}=(0,0,1)$, calculated for TE polarization and $l=7$ via the RSs summation Eq.\,(9) ($F_7(\omega)$, solid line) and by using the analytic form of the dyadic GF ($F^{\rm a}_7(\omega)$, dotted line). (b) as (a) but for TM polarization. (c) Errors $F^{\rm a}_7(\omega)-F_7(\omega)$ for TE and TM polarization using a summation cutoff $\omega_{\rm max}=40$.
}\label{fig:SM5}
\end{figure}

\begin{figure}[t!]
\includegraphics*[width=\columnwidth]{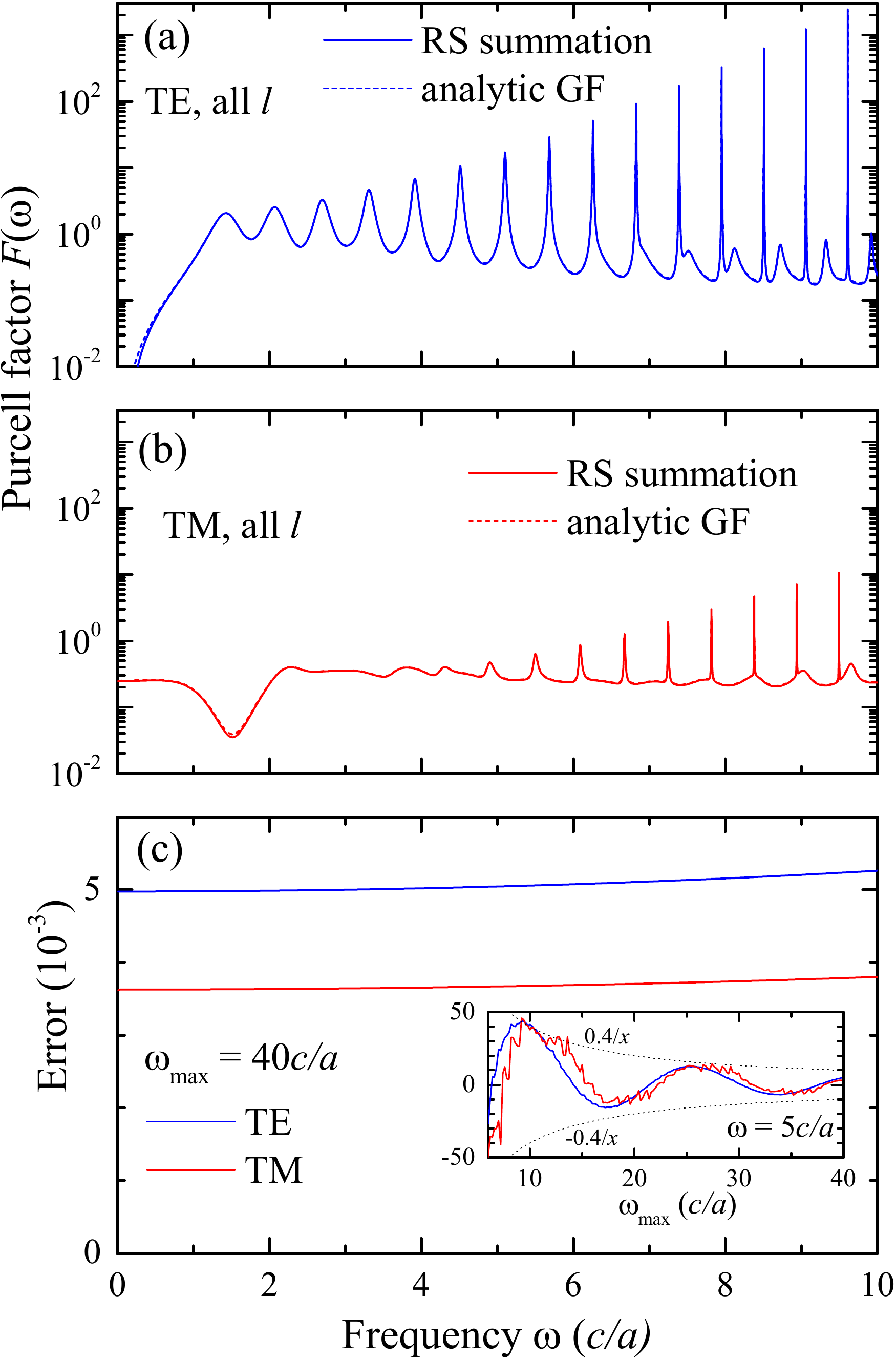}
\caption{(a) Full Purcell factor of a dielectric sphere in vacuum, with permittivity $\varepsilon =4$ and radius $a$, for a point dipole placed at $|\br_d|=0.9a$ with direction ${\bf e}=(0,0,1)$, calculated for TE polarization via the RSs summation Eq.\,(9) ($F(\omega)$, solid line) and by using the analytic form of the dyadic GF ($F^{\rm a}(\omega)$, dotted line). (b) as (a) but for TM polarization. (c) Errors $F^{\rm a}(\omega)-F(\omega)$ for TE and TM polarization using a summation cutoff $\omega_{\rm max}=40c/a$. The inset shows the errors at $\omega=5c/a$ as function of $\omega_{\rm max}$, with the convergence illustrated by the dotted lines showing $\pm0.4c/(\omega_{\rm max}a)$.
}\label{fig:SM6}
\end{figure}

The collective mode volumes of several RSs with $l=7$, calculated using Eqs.~(\ref{V-TE})--(\ref{E-TETM}), are shown in Fig.~1(a) of the main text for TE polarization and in Figs.~\ref{fig:SM3}(a) and \ref{fig:SM4}(a) for TM polarization and two different directions of the dipole. We note that the fundamental $n=1$ WGMs in TE and TM polarizations, which have quite similar Q-factors of the order of 100, have very different mode volumes for a given direction of the dipole. Indeed, for an azimuthal dipole direction  ${\bf e}=(0,0,1)$ the effective volume of the TE mode is much smaller than the one of the TM mode. This is because the electric field in TM polarization is mostly in radial direction, with only a small azimuthal component. For a radial direction of the dipole ${\bf e}=(1,0,0)$ instead, the TM mode has a much smaller mode volume, comparable to that of the TE mode for ${\bf e}=(0,0,1)$, as seen by comparing Fig.\,1(a) and \Fig{fig:SM4}(a). The partial PFs due to all $l=7$ modes within the spectral range up to $\omega_n a/c\sim 40$ are shown separately, in Fig.\,1(b) for TE and in Figs.\,\ref{fig:SM3}(b) and \ref{fig:SM4}(b) for TM polarization. These figures demonstrate the strong dependence of the PF on the dipole orientation, as discussed above. Summing over all different $l$ components and averaging over all possible directions of the dipole, we obtain the full PF for this system which is demonstrated in Fig.~2 of the main text.

\section{Verification of the exact formula for the Purcell factor}

To verify the exact formula Eq.\,(9) for the PF we compare both the partial PF $F_l(\omega)$ for a given $l$, and the full PF $F(\omega)$ (i.e. the sum over all $l$) with those obtained via Eq.\,(5) using a direct evaluation of the dyadic GF, resulting in $F^{\rm a}_l(\omega)$ and $F^{\rm a}(\omega)$, respectively. The analytic form of the dyadic GF for a dielectric sphere in vacuum is known in the literature~\cite{Chew} and can be represented in terms of the linearly independent solutions of the second-order differential equation, given by spherical Bessel and Hankel functions. We use this analytic form both in TE and TM polarizations, for the same dielectric sphere and the same position of the dipole as considered in Figs.\,1 and \ref{fig:SM3}. We find excellent agreement, limited only by the finite number of RSs taken into account in the sums. As an example, we compare $F^{\rm a}_7(\omega)$ with $F_7(\omega)$ in \Fig{fig:SM5}\,(a) and (b). The error $F^{\rm a}_7(\omega)-F_7(\omega)$ given in \Fig{fig:SM5}\,(c) shows that the analytic result is about $10^{-4}$ higher. This deviation is due to the limited number of RSs used to calculate $F_7$ via Eq.\,(9) -- we took into account all RSs with $|\omega_n|<\omega_{\rm max}=40c/a$, as in the main text. The missing contributions of the higher frequency RSs yield a underestimation of the PF, which is increasing with frequency as one approaches the frequencies of the missing RSs.

Summing over all partial PFs with $l<\omega_{\rm max}a/c$, we obtain the full PF, both in TE and TM polarizations for the same dipole, see \Fig{fig:SM6}\,(a) and (b). Its error $F^{\rm a}(\omega)-F(\omega)$ shown in \Fig{fig:SM6}\,(c) is again positive and slightly increasing with frequency. It is larger than the partial error shown in \Fig{fig:SM6}\,(c) due to the accumulated errors from different $l$.
These errors are oscillating and decaying in magnitude with increasing cutoff frequency $\omega_{\rm max}$, as demonstrated in the inset of \Fig{fig:SM6}\,(c). A convergence faster than $\omega^{-1}_{\rm max}$ is observed, as indicated by the dotted lines.



\section{Application of exact normalization to modes calculated with numerical solvers}
\label{Sec-Comsol}

\newcommand{\Dmesh}{\Delta_{\rm m}}
\newcommand{\Dgrid}{\Delta_{\rm g}}

We have shown in \Fig{fig:SM1} and \Fig{fig:SM2} the relative error of the LK normalization, demonstrating sizeable errors close to the system (of around 10\% for the examples shown), and diverging errors in the limit to infinite radius $R$. We can see that, excluding the leaky modes which have $Q<3$, the divergence for large radii only becomes apparent for $R/a>10$, a regime that in a numerical simulation would not likely be explored due to the large required computational domain. The errors at small radii are therefore typically the more significant limitation of the LK normalization for simulations of high Q modes.

It is worth at this point to give an intuitive picture of the difference between the LK and the exact normalization, which we mentioned in the main text. If we consider a plane wave propagating with wavevector $\bk$ in vacuum, we can write its phase as $\bk\cdot \br= r(\omega/c)\cos\alpha$, where $r$ is the magnitude of $\br$ and $\alpha$ is the angle between the propagation direction and the  direction of $\br$. Evaluating the surface term of the normalization on a small part of a spherical surface, we obtain that the surface term in Eq.\,(8) is a factor $\cos \alpha$ smaller than in Eq.\,(4). This gives us a physical picture of the difference between the two normalizations. The LK normalization assumes a propagation normal to the surface of integration, while the exact normalization takes into account the actual propagation direction. To determine the actual propagation direction, spatial derivatives of the fields are required, and are therefore present in Eq.\,(8).

We now can also understand why the LK normalization would be valid for $s$-waves and a spherical surface around the center -- in this case the propagation direction would be radial. However, pure $s$-wave (such as $l=0$ modes of a sphere) do not exist in electrodynamics of finite systems, as already mentioned in Sec.\,\ref{Sec-Krist}. We note that Kristensen {\it et al.} have recently published an article \cite{KristensenARX15} in which they show in Fig.\,2 the LK normalization for a geometry with an effectively $s$-wave mode, using a two-dimensional problem with translation invariance along $z$ and electric field along $z$. This is a well chosen example to show that the LK normalization can have small errors. The second example shown is an axially symmetric geometry of a gold nanorod dimer, and a dipolar mode is shown having the emitted field dominantly polarized along the axial direction and thus again a dominant $s$-character, leading to reduced errors of the LK normalization.

Using this insight we can understand the resulting errors of the LK normalization. At small distances from the system, the propagation direction will significantly differ from being radial, simply due to geometrical conditions. This is the reason for the significant errors at small distances seen in \Fig{fig:SM1} and \Fig{fig:SM2}. Going to large distances, the angle $2\alpha$, at which the system is seen from the observation point, will scale as $1/R$, so that the error, $1-\cos\alpha$, scales as $1/R^2$. Note that this is consistent with the analytical result for the Mie modes in \Eq{Vratio1} and also in \Onlinecite{KristensenARX15} Eq.\,(26). For resonant states with finite loss, this leads to the divergence at large distances $\propto \exp(2i\omega_n R/c)/R^2$, where the exponential increase of the field dominates the $R^{-2}$ decay of the LK normalization error.

Importantly, this shows that even if we can neglect the error due to the exponential divergence, for example by considering high Q modes, the error of the surface term in Eq.\,(4) is still approximately $L^2/(2R)^2$, where $L$ is the system size. In order to provide an error of 1\%, which could be considered sufficient, we still need a simulation size which is about five times the system size. Simulating such large areas is computationally costly, specifically in three dimensions, where a simulation volume of about 100 times the system volume would be required. Notably, the LK normalization requires a spherical normalization volume with the system in the center, in order to provide this $R^{-2}$ scaling of the propagation direction error, while for other shapes it is not converging to the correct normalization even for modes of infinite Q.

The exact normalization instead can be used for any volume enclosing the system, with arbitrary volume shapes. Therefore the exact normalization is also advantageous for numerical evaluation, as it does not require an extended simulation domain.

In order to exemplify this discussion and explicitly show the applicability of the exact normalization to modes determined with numerical solvers, we have calculated modes of a dielectric cylinder using the FEM solver Comsol. The cylinder has a height $h$ equal to its diameter $2a$, and a refractive index of 3. We used the axi-symmetric eigenmode solver of Comsol, and an angular quantum number of $m=1$. The cylinder is embedded in vacuum and the simulation area is enclosed by a spherical PML at a radius of $5\lambda+2\sqrt{2}a$, where $\lambda$ is the target vacuum mode wavelength $\lambda=2a$. The PML thickness was $\lambda$. The mesh was determined by a single mesh parameter $\Dmesh$, from which the mesh was created using a free triangular mesh with a minimum size of $\Dmesh/3$ in the cylinder, $\Dmesh$ in vacuum and the PML, and a maximum size twice the minimum size in all regions.  It is important to note that we used a large simulation domain to show the $R$ dependence of the normalization, which leads to large numerical complexity. Furthermore, surface modes (whispering gallery modes) are forming at the vacuum to PML interface due to the non-ideality of the PML, which are constituting most modes found by the eigenmode solver of Comsol when using such a large simulation domain, making it cumbersome to find the non-spurious modes of the system.

To evaluate the normalizations integrals, we exported the mode field into a square grid of $\Dgrid$ pitch. We calculated the first and second derivatives using a three-point differential scheme with an error scaling as ${\cal O}(\Dgrid^2)$. The surface integrals were evaluated by dividing the half circle in cylindrical coordinates representing the surface into 1025 linear segments. More details on the computational method and the code used can be found in the data DOI of this work and on langsrv.astro.cf.ac.uk/Normalization.

\begin{figure}[t!]
	\includegraphics*[width=\columnwidth]{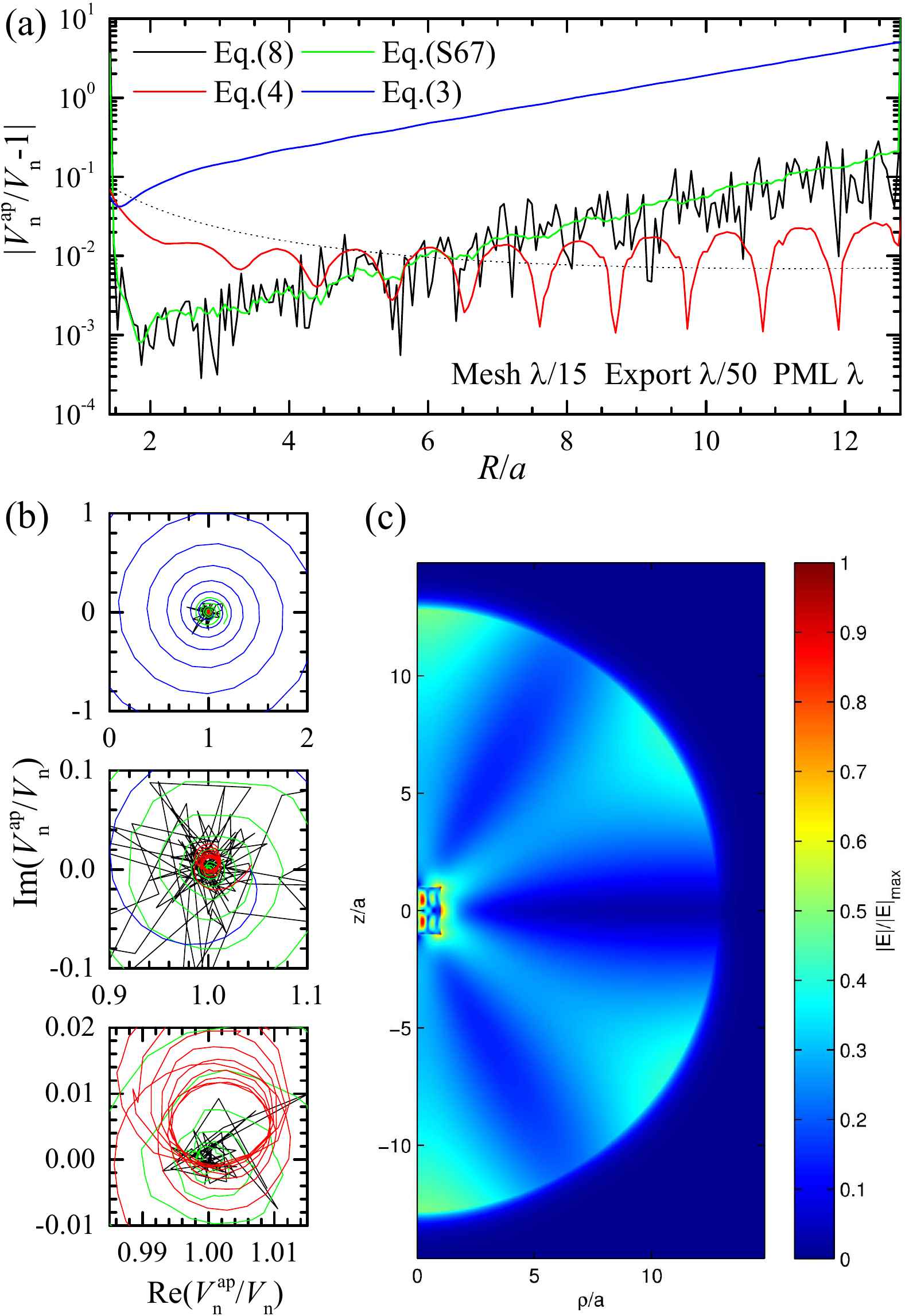}
	\caption{Normalization of a RS of angular quantum number $m=1$ of a cylinder of radius $a$ and height $2a$, calculated using ComSol, using  $\Dmesh=\lambda/15$, and an export mesh $\Dgrid=\lambda/50$. (a) Relative error of the approximate mode volume $|V_n^{\rm ap}/V_n-1|$ as function of the radius $R$ of the sphere of integration. $V_n^{\rm ap}$ is calculated without surface term Eq.\,(3), or the LK surface term Eq.\,(4). Additionally the results of the correct surface term using first and second derivatives Eq.\,(8) and only first derivatives \Eq{norm-disp4}. The dashed line indicates a term $\propto |\exp(2i\omega_n R/c)/R^2|$. (b) Relative complex approximate mode volume $V_n^{\rm ap}/V_n$ for $R/a$ varying from 1.5 to 12.3. Three different zooms are shown. Calculated points are connected by straight lines. For the highest zoom the results of Eq.\,(8) are shown only for $R/a<5$.  (c) Mode field amplitude $|E_n(\br)|$ in cylindrical coordinates $(\rho,z)$.
	}\label{fig:CSL15}
\end{figure}

We show here the results for a mode with a frequency of $\omega_n a/c = 2.943508477-0.1724037812i$ having a $Q$ of about 8.5. For illustration, the electric field distribution in the cylindrical coordinate plane $(\rho,z)$ is shown in \Fig{fig:CSL15}(c). The far-field emission pattern has three nodes showing its non-$s$-wave character. The effect of the exponential divergence of the mode towards larger radius is evident, as the $R^{-1}$ decay due to the three-dimensional emission is superseded by the exponential growth of $|\exp(i\omega_n R/c)|$.

To determine the normalization we evaluate Eq.\,(8) as function of $R$. We find numerical noise which increases with increasing $R$, proportional to the surface term. We therefore choose as mode normalization $V_n$ the value for which the surface integral is small, at about $R/a=2$.  The resulting $R$-dependent normalization errors are given in \Fig{fig:CSL15}(a). Let us first discuss the result of Eq.\,(8). We can see a value which is randomly fluctuating around zero [see also the complex normalized mode volume in \Fig{fig:CSL15}(b)], showing that the deviation is due to numerical errors. Comparing this error with the error of the volume only normalization Eq.\,(3), we see that it is about 1-2 orders of magnitude smaller, indicating that the relative error of the surface integral evaluation is about 1-10\% in this case. Interestingly, using the first-derivative expression for the exact normalization, \Eq{norm-disp4}, this random error is converted into a systematic error of similar magnitude, but spiralling in the complex plane.
We have found that choosing different $\Dgrid$ changes these numerical errors, and we have chosen the ratio between $\Dgrid$ and $\Dmesh$ to provide the smallest errors  within a factor of two for both exact normalization formulations. We noted that for \Eq{norm-disp4} the optimum  $\Dgrid$ was about three times smaller than for Eq.\,(8).

The error of the correct surface integral is dominated by the error in determining the local propagation direction, i.e. by the spatial derivatives of the fields. To minimize the resulting error of the normalization, it is therefore best to evaluate the normalization using a volume with a small surface term, as we have done. Too close to the system, strong field gradients and spatial variations of the mesh can give rise to additional numerical errors, as visible in \Fig{fig:CSL15}(a) for $R<2a$.

We find the error of Eq.\,(4) to be significant for small $R$, similar to what observed for the Mie modes in \Fig{fig:SM1} and explained as the failure of the radial propagation assumption of Eq.\,(4). With increasing $R$ the value oscillates, due to a rotation of the error in the complex plane as shown in \Fig{fig:CSL15}(b), according to the scaling $\exp(2i\omega_n R/c)/R^2$ shown by dashed line in \Fig{fig:CSL15}(a).  Additional to the oscillation, we also observe a slow drift of the center position, which we discuss later on.
\begin{figure}[t!]
	\includegraphics*[width=\columnwidth]{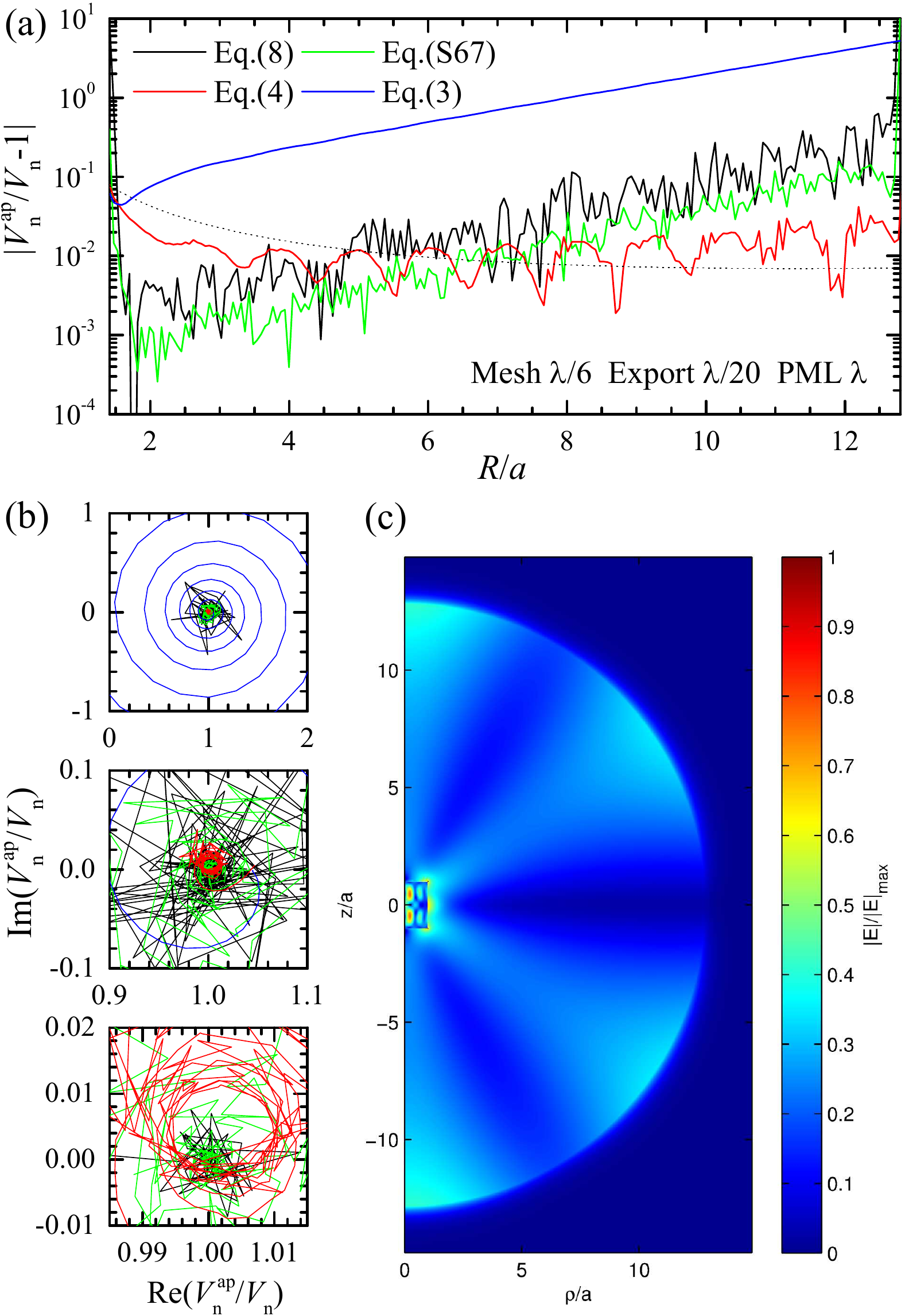}
	\caption{As. \Fig{fig:CSL15}, but using a mesh of $\lambda/6$, and an export mesh $\lambda/20$.
	}\label{fig:CSL6}
\end{figure}

We now compare these result with the ones for a 2.5 times coarser mesh ($\Dmesh=\lambda/6$ and $\Dgrid=\lambda/20$) shown in \Fig{fig:CSL6}, which leads to a very small relative change of $\omega_n$ of $-1.2\cdot10^{-6}$ for the real part and $-7.8\cdot10^{-5}$ for the imaginary part. This mesh size is advised in typical Comsol examples. We find that the results are generally similar, but show higher random numerical errors. Notably, we find that the first derivative formulation \Eq{norm-disp4} shows now an error about five times lower than the one of the second derivative formulation Eq.\,(8), and having a similar magnitude as for the  finer mesh shown before. This illustrates an advantage of the first-derivative normalization for coarser grid calculations having larger numerical errors.

\begin{figure}[t!]
	\includegraphics*[width=\columnwidth]{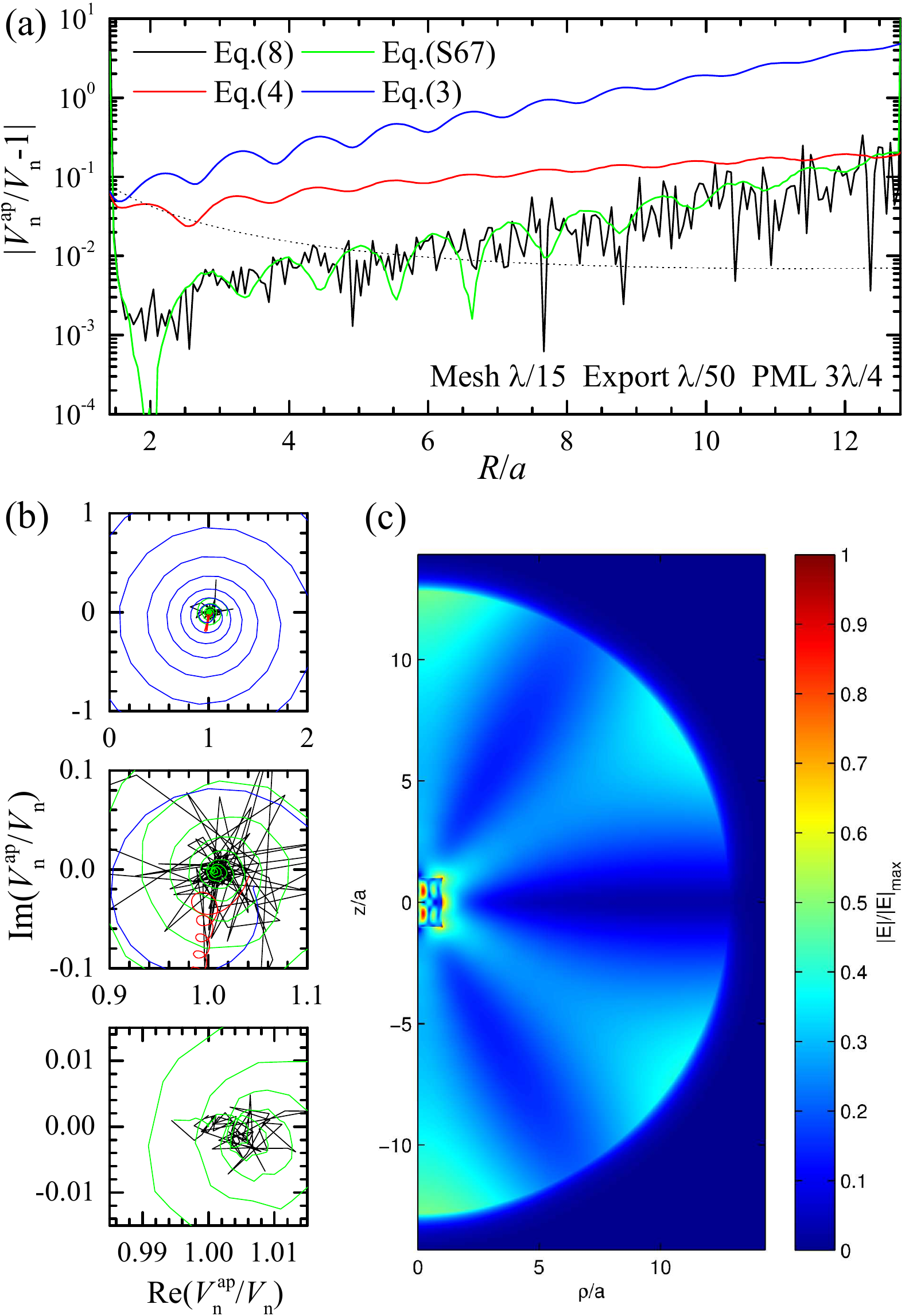}
	\caption{As. \Fig{fig:CSL15}, but using a PML thickness of $3\lambda/4$.
	}\label{fig:CSL15P0p75}
\end{figure}

We now look at the effects of the non-ideality of the PML, producing reflected waves. Such waves are incoming, and thus have the opposite propagation direction to the one assumed in Eq.\,(4). We used the fine mesh of \Fig{fig:CSL15}, but a PML thickness reduced from $\lambda$ to $3\lambda/4$. This leads to a relative change of $\omega_n$ of $-3.5\cdot10^{-3}$ for the real part and $-8.2\cdot10^{-3}$ for the imaginary part. While the presence of the reflected wave is hardly visible in the mode amplitude shown in \Fig{fig:CSL15P0p75}(c), the error of Eq.\,(3) in \Fig{fig:CSL15P0p75}(a) is clearly showing oscillations due to the interference of outgoing and incoming waves. Their contrast is increasing with decreasing $R$ due to the exponential growth of the reflected incoming field with decreasing  $R$, similar to the growth with increasing  $R$ of the outgoing field. Evaluating the total field growth during the propagation to the PML and back to the center due to Im$(\omega_n)$ for the present case, we find a factor of about 15. Using large simulation volumes as done here results in such large factors, which give rise to artefacts even for a low reflectivity of the PML. Again for the present case, we see a contrast of about 20\% close to the center, corresponding to the incoming field amplitude of about 10\% of the outgoing one, and thus to a PML amplitude reflectivity of about 0.7\%, or intensity reflectivity of about $5\times 10^{-5}$. Very small residual reflectivities of the PML can thus lead to large contributions of incoming waves, specifically for modes of low $Q$.

Looking at the error of the LK normalization in \Fig{fig:CSL15P0p75}(a), we see the significant impact of the incoming waves. The behavior in the complex plane in \Fig{fig:CSL15P0p75}(b) is instructive -- additionally to the spiraling the normalization is drifting in phase approximately linearly with $R$ by up to 0.2 radians.

\begin{figure}[t!]
	\includegraphics*[width=\columnwidth]{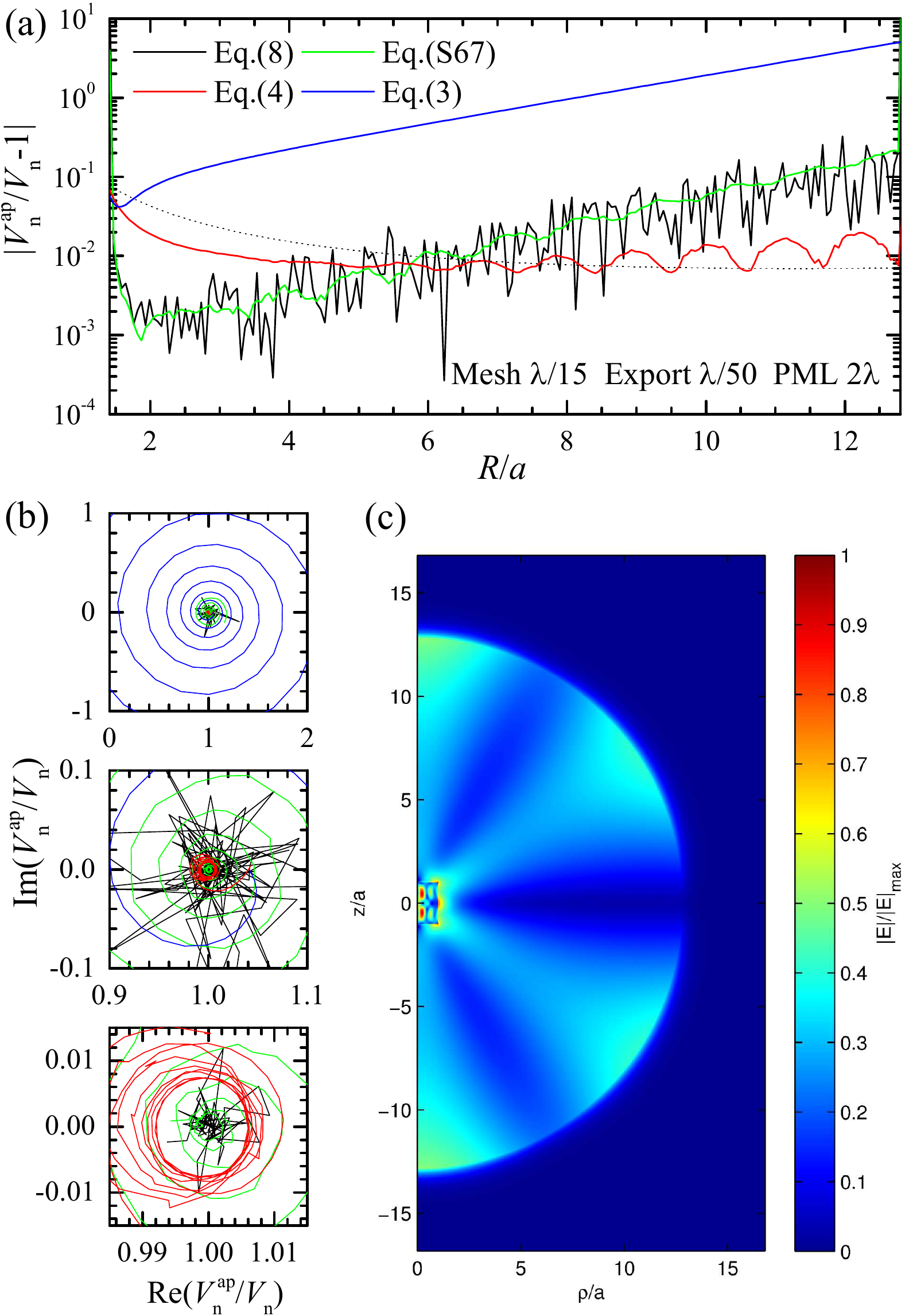}
	\caption{As. \Fig{fig:CSL15}, but using a PML thickness of $2\lambda$.
	}\label{fig:CSL15P2}
\end{figure}

The drift of the LK normalization seen in \Fig{fig:CSL15} is therefore attributed to a small reflection of the PML, which can be also seen as weak oscillation of the volume normalization. Increasing the thickness of the PML to $2\lambda$, shown in \Fig{fig:CSL15P2}, this drift is reduced, and the center of rotation of the LK normalization is found closer to the exact normalization. The relative change of $\omega_n$ due to this increased PML thickness is $2.1\times10^{-4}$ for the real part and $6\times10^{-4}$ for the imaginary part.

Interestingly, the exact normalization is hardly influenced by such reflections, which is understandable since it takes into account the propagation direction. To compare the normalization of the modes for the different simulations, we use the volume integral Eq.\,(3) evaluated for the smallest possible radius of the sphere fully including the cylinder,   $R=\sqrt{2}a$. This integral is proportional to the electric field squared and consequently to the inverse mode volume. For the exact normalization of the mode, we find the value of the volume integral to be $1.05735-0.00126i$ for \Fig{fig:CSL15},  $1.06158-0.00108i$ for \Fig{fig:CSL6}, $1.06469-0.01038i$ for \Fig{fig:CSL15P0p75}, and $1.05704-0.00202i$ for \Fig{fig:CSL15P2}. This demonstrates that the exact normalization is stable to 0.7\% for all these cases. At the same time the LK normalization has minimum errors above 4\% in the case of \Fig{fig:CSL15P0p75}.

To summarize, we have presented an example of the applicability of the exact normalization to numerically determined modes, which demonstrates the following. Firstly, as the exact normalization can be evaluated over any volume containing the system,  this volume can be chosen close to the system to have a small surface term and therefore smaller errors, and not requiring an extension of the simulation domain. Secondly, taking proper account of the propagation direction makes the exact normalization less susceptible to incoming waves, such as those propagating from non-ideal PMLs, so that the exact normalization is robust against such errors in numerical simulations.

Considering the LK normalization instead, we emphasize that it assumes that the field at the surface is propagating normal to the surface of the normalization volume. It therefore does not determine the propagation direction from the field gradients, and is consequently having smaller numerical errors. However, this assumption creates systematic errors which depend on the specific mode analyzed and the surface geometry used. For the widely used spherical volume, the resulting error scales $\propto \exp(2i\omega_n R/c)/R^2$, where the $1/R^2$ factor simply comes from the angular size of the system seen from the surface of integration. If the mode frequency $\omega_n$ is real, the error is converging to zero at $R\to\infty$. For any leaky mode the finite imaginary part of $\omega_n$ leads to a divergence of the error at $R\to\infty$. For modes with a sufficiently large Q factor, this divergence however is seen only for $R$ larger than any numerically treatable domains. Even in this case, a significant reduction of the errors due to non-normal propagation requires large values of $R$, as demonstrated by Figs.\,\ref{fig:SM1} and \ref{fig:CSL15}--\ref{fig:CSL15P2}. Therefore in practical terms, the most problematic feature of the LK normalization for high Q modes is not the divergence at $R\to\infty$, but the requirement to use simulation sizes much larger than the system size, in order to approximately achieve the normal incidence condition.
